\documentclass{IEEEtran}
\usepackage[cmex10]{amsmath}
\usepackage{ulem}
\usepackage{cite}
\usepackage{graphicx}
\usepackage{amsmath}
\usepackage{array}
\usepackage{mdwmath}
\usepackage{amssymb}
\usepackage{mdwtab}
\usepackage{subfigure}
\usepackage{url}
\usepackage{stfloats}
\usepackage{amsmath,amsthm}
\usepackage{verbatim}
\usepackage{threeparttable}
\usepackage{color}
\usepackage{ulem}
\usepackage{bm}

\newtheorem{theorem}{Theorem}

\newtheorem{lemma}{Lemma}
\newtheorem{corollary}{Corollary}

\def\BibTeX{{\rm B\kern-.05em{\sc i\kern-.025em b}\kern-.08em
    T\kern-.1667em\lower.7ex\hbox{E}\kern-.125emX}}
\usepackage{multirow}
\usepackage{booktabs}

\usepackage{array}
\newcolumntype{L}[1]{>{\raggedright\let\newline\\\arraybackslash\hspace{0pt}}m{#1}}
\newcolumntype{C}[1]{>{\centering\let\newline\\\arraybackslash\hspace{0pt}}m{#1}}
\newcolumntype{R}[1]{>{\raggedleft\let\newline\\\arraybackslash\hspace{0pt}}m{#1}}

\def\BibTeX{{\rm B\kern-.05em{\sc i\kern-.025em b}\kern-.08em
    T\kern-.1667em\lower.7ex\hbox{E}\kern-.125emX}}
\begin{document}

\title{Secret Key Generation for IRS-Assisted  Multi-Antenna Systems: A Machine Learning-Based Approach
}

\author{Chen Chen, \textit{Member, IEEE},  Junqing~Zhang, \textit{Member, IEEE}, Tianyu~Lu, \textit{Graduate Student Member, IEEE}, \\ Magnus~Sandell, \textit{Senior Member, IEEE}, Liquan~Chen, \textit{Senior Member, IEEE}

\thanks{Manuscript received xxx; revised xxx; accepted xxx. Date of publication xxx; date of current version xxx. Part of this paper has been accepted by IEEE ICC 2023 \cite{chen2023machine}. The work was in part supported by UK EPSRC under grant ID EP/V027697/1 and in part by National Key Research and Development Program of China under grant ID 2020YFE0200600. The review of this paper was coordinated by xxx. 	\textit{(Corresponding author: Junqing Zhang.)}}

\thanks{C. Chen and J. Zhang are with the Department of Electrical Engineering and Electronics, the University of Liverpool, Liverpool, L69 3GJ, UK. (email: c.chen77@liverpool.ac.uk; Junqing.Zhang@liverpool.ac.uk)

T.~Lu and L.~Chen are with the School of Cyber Science and Engineering, Southeast University, Nanjing, 210096, China. (email: effronlu@seu.edu.cn; lqchen@seu.edu.cn)

M.~Sandell is  with Bristol Research and Innovation Laboratory, Toshiba Research Europe Ltd., Bristol, BS1 4ND, UK. (email: magnus.sandell@toshiba-bril.com)}

\thanks{Color versions of one or more of the figures in this paper are available online at http://ieeexplore.ieee.org.}
\thanks{Digital Object Identifier xxx}	
}

\maketitle

\begin{abstract}
Physical-layer key generation (PKG) based on wireless channels is a lightweight technique to establish secure keys  between legitimate communication nodes. Recently, intelligent reflecting surfaces (IRSs) have been leveraged to enhance the performance of PKG in terms of secret key rate (SKR), as it can reconfigure the wireless propagation  environment and introduce more channel randomness. In this paper, we investigate an IRS-assisted PKG system, taking into account the channel spatial correlation at both the base station (BS) and the IRS. Based on the considered system model, the closed-form expression of SKR is  derived analytically considering correlated eavesdropping channels. Aiming to maximize the SKR, a joint design problem of the BS's precoding matrix and the IRS’s phase shift vector is formulated. To address this high-dimensional non-convex optimization problem, we propose a novel unsupervised deep neural network (DNN)-based algorithm with a simple structure. Different from most previous works that adopt iterative optimization to solve the problem, the proposed DNN-based algorithm directly obtains the BS precoding and IRS phase shifts as the output of the DNN. Simulation results reveal that the proposed DNN-based algorithm outperforms the benchmark methods with regard to SKR. 
\end{abstract}

\begin{IEEEkeywords}
Physical-layer key generation, intelligent reflecting surfaces, deep neural network
\end{IEEEkeywords}

\section{Introduction}
Along with the surge in data traffic envisioned by sixth generation (6G) communication, data security risks emerge due to the broadcast nature of wireless transmissions~\cite{Magzine1, nguyen2021security}. Conventional cryptographic  approaches rely on either asymmetric or symmetric encryption for achieving data confidentiality. Asymmetric cryptography, also known as public key cryptography (PKC), can be used for data encryption and key distribution. The latter can share keys for symmetric encryption, e.g., advanced encryption standard (AES).
However, PKC is computationally complicated and requires public key infrastructure for the distribution of keys. 
Therefore, conventional cryptographic  approaches are challenging to be applied to mobile communication networks and resource-constrained Internet of Things (IoT).

In this context, physical-layer key generation (PKG) has been proposed as a lightweight technique that does not rely on conventional cryptographic  schemes and is information-theoretically secure~\cite{Access1}. PKG utilizes the reciprocity of wireless channels during the channel coherent time to generate symmetric keys from bidirectional channel probings. The spatial decorrelation and channel randomness ensure the security of the generated keys. More specifically, 
spatial decorrelation ensures that eavesdroppers cannot generate the same keys as legitimate communicating nodes, and channel randomness prevents the generated keys from dictionary attacks.

Although the effectiveness of PKG has been demonstrated both theoretically and experimentally~\cite{waqas2018social, bakcsi2019secret, premnath2012secret, zhang2018channel, chen2022sample},  PKG may face serious challenges in harsh propagation environments. The transmission link between two legitimate nodes may experience non-line-of-sight (NLoS) propagation conditions due to the  effects  of blockages. In this case,  the channel estimation will suffer from a low signal-to-noise ratio (SNR), resulting in a high bit disagreement rate (BDR) and  low  secret key rate (SKR).  Besides, in a quasi-static environment that lacks randomness, the achievable SKR is limited due to low channel entropy \cite{Access1, aldaghri2020physical}. 

To address these challenges, a new paradigm, called intelligent reflecting surface (IRS), offers attractive solutions.
IRS consists of low-cost passive reflecting elements that can dynamically adjust their amplitudes and/or phases to reconfigure the  wireless propagation environments in an energy-efficient manner~\cite{bjornson2022reconfigurable, wu2021intelligent}.  The reflection channels provided by IRS are capable of enhancing the received signals at legitimate users and introducing rich channel randomness. 
There have been several works exploring the application of IRS in PKG \cite{CL1, staat2021intelligent, TVT1, TIFS2}.
The work in~\cite{CL1} improved the SKR of an IRS-aided single-antenna system by randomly configuring the IRS's phase shifts, in which closed-form expressions of the lower and upper bounds of the SKR were provided. 
In \cite{staat2021intelligent}, a practical prototype of IRS-assisted PKG system was implemented using commodity Wi-Fi transceivers, which demonstrated that the randomly configured IRS can boost the key generation rate while ensuring the randomness of the generated keys. To further improve the SKR, some works developed optimization algorithms to configure the IRS’s phase shifts.
In~\cite{TVT1}, the IRS phase shifts were optimized to maximize the minimum achievable SKR. Successive convex approximation was adopted to address the non-convex optimization objective. In~\cite{TIFS2}, the authors considered a multi-user scenario and designed IRS phase shifts to maximize the sum SKR. 
 Nevertheless, these works only considered the optimization of IRS phase shifts in a single-antenna scenario, and thus cannot be applied to multi-antenna systems that require a joint design of multi-antenna precoding and IRS phase shifts.  

Multi-antenna technique has been a key enabler of fifth
generation (5G) and beyond wireless communication networks \cite{zhang2020prospective, chen2022deployment}. Therefore, securing IRS-assisted multi-antenna transmissions using PKG becomes an important topic. Compared to IRS-assisted PKG efforts in single-antenna scenarios, there is limited work investigating PKG in  IRS-assisted multi-antenna systems~\cite{hu2022reconfigurable, lu2022joint}. In \cite{hu2022reconfigurable}, the authors maximized the minimum SKR by jointly optimizing the transmit and the reflective beamforming through a block successive upper-bound minimization-based algorithm. The unit pilot length was adopted for both uplink and downlink channel probings, which simplified the analysis but limited the SKR. In \cite{lu2022joint}, the BS precoding and IRS phase shift matrices were jointly designed using
fine-grained channel estimation. However, the analysis of correlated eavesdropping channels was overlooked.
Generally, the optimal configuration of IRS is a complex non-convex optimization problem.  This problem is exacerbated in multi-antenna systems that require a joint optimization of transmit precoding and reflection beamforming. The existing works rely on iterative optimization algorithms with a high computational complexity, which is difficult to be deployed in practice.

Recently, machine learning has emerged as one of the key techniques to address mathematically intractable non-convex optimization problems~\cite{Survey1, Magzine2}. There have been research efforts~\cite{Access2, JSAC1, zhang2021deep, chen2023distributed} leveraging machine learning to optimize IRS reflecting beamforming with the aim of maximizing downlink transmission rates. A recent work~\cite{jiao2021machine} proposed a machine learning-based adaptive quantization method to balance key generation rate and BDR in an IRS-aided single-antenna system, where the IRS phase shifts were configured using iterative optimization. Up to now, the employment of machine learning to optimize  IRS phase shifts in an IRS-aided PKG system has not been investigated.

To the best of our knowledge, this is the first attempt to exploit machine learning in jointly optimizing the transmit precoding and IRS reflection beamforming for maximizing the SKR in an IRS-assisted multi-antenna system.  Our
major contributions are summarized as follows: 
\begin{itemize}
	\item We propose a new PKG framework for an IRS-assisted multi-antenna system and derive the closed-form expression of SKR considering correlated eavesdropping channels and spatial correlation among BS antennas and IRS elements.
	\item We formulate the optimization problems of SKR maximization in the absence and presence of eavesdropper channel statistics, respectively.
Then water-filling algorithm-based baseline solutions are developed to jointly design the BS's precoding matrix and the IRS's phase shift vector.
	\item We novelly propose to obtain the optimal configuration of BS precoding and IRS phase shifts by using unsupervised deep neural networks (DNNs), referred to as ``PKG-Net''. The proposed PKG-Net can deal with different transmit power levels and channel statistics parameters. Simulations demonstrate that the proposed PKG-Net can achieve a higher SKR than other benchmark methods. Compared with the  water-filling algorithm-based baseline solution, the proposed PKG-Net has a significantly lower computational complexity. 
\end{itemize}
Different from our previous work \cite{chen2023machine}, where the eavesdropping channels are assumed to be uncorrelated with the legitimate channels, in this paper, we considerably extend the system to a more general scenario in the presence  of correlated eavesdropping channels. Moreover, the DNN developed in \cite{chen2023machine} only has generalization ability to different user locations. 
In this paper, we further extend the generalization ability of PKG-Net to different transmission powers and channel statistics.

The rest of this paper is organized as follows. In Section \ref{sec:systen_model}, we present the system model. In Section \ref{sec:secret_key rate}, we derive the expression of SKR. The problem formulation is presented in Section \ref{sec:problem_formulation}. Then we propose a water-filling algorithm-based
baseline solution in Section \ref{sec:baseline}. A machine-learning-based method is developed in Section \ref{sec:machine_learning}.
 Our simulation results and analysis are given in Section \ref{sec:simulation}. Finally, Section \ref{sec:conslusions} concludes this paper. 

Notations: In this paper, scalars are represented by italic letters, vectors
by boldface lower-case letters,
and matrices  by boldface uppercase letters. $\mathbf{V}^{T}$, $\mathbf{V}^{H}$ and $\mathbf{V}^{*}$ are the transpose, conjugate transpose
and conjugate of a matrix $\mathbf{V}$, respectively.
$\mathbb{E}\{\cdot\}$ is
the statistical expectation. $\mod(\cdot)$ and $\lfloor\cdot\rfloor$ denote modulus operator and  floor function, respectively. $[\mathbf{V}]_{i,j}$ denotes the $(i,j)$-th element of a matrix $\mathbf{V}$. $\mathcal{CN}(\mu,\sigma^2)$ represents a circularly symmetric complex Gaussian  distribution with mean $\mu$ and variance $\sigma^2$.  $\text{diag}(\mathbf{v})$ is a diagonal matrix with the entries of $\mathbf{v}$ on its main diagonal and $\text{vec}(\mathbf{V})$ is the vectorization of a matrix $\mathbf{V}$. 
$\mathbb{C}^{A\times B}$ represents the sapce of a complex matrix with size $A\times B$.
$||\cdot||_F$ denotes the Frobenius norm. $\odot$ and $\otimes$ are the Hadamard product and Kronecker product, respectively. 

\section{System Model}
\label{sec:systen_model}
\subsection{System Overview}
\begin{figure}[!t]
\centerline{\includegraphics[width=3.4in]{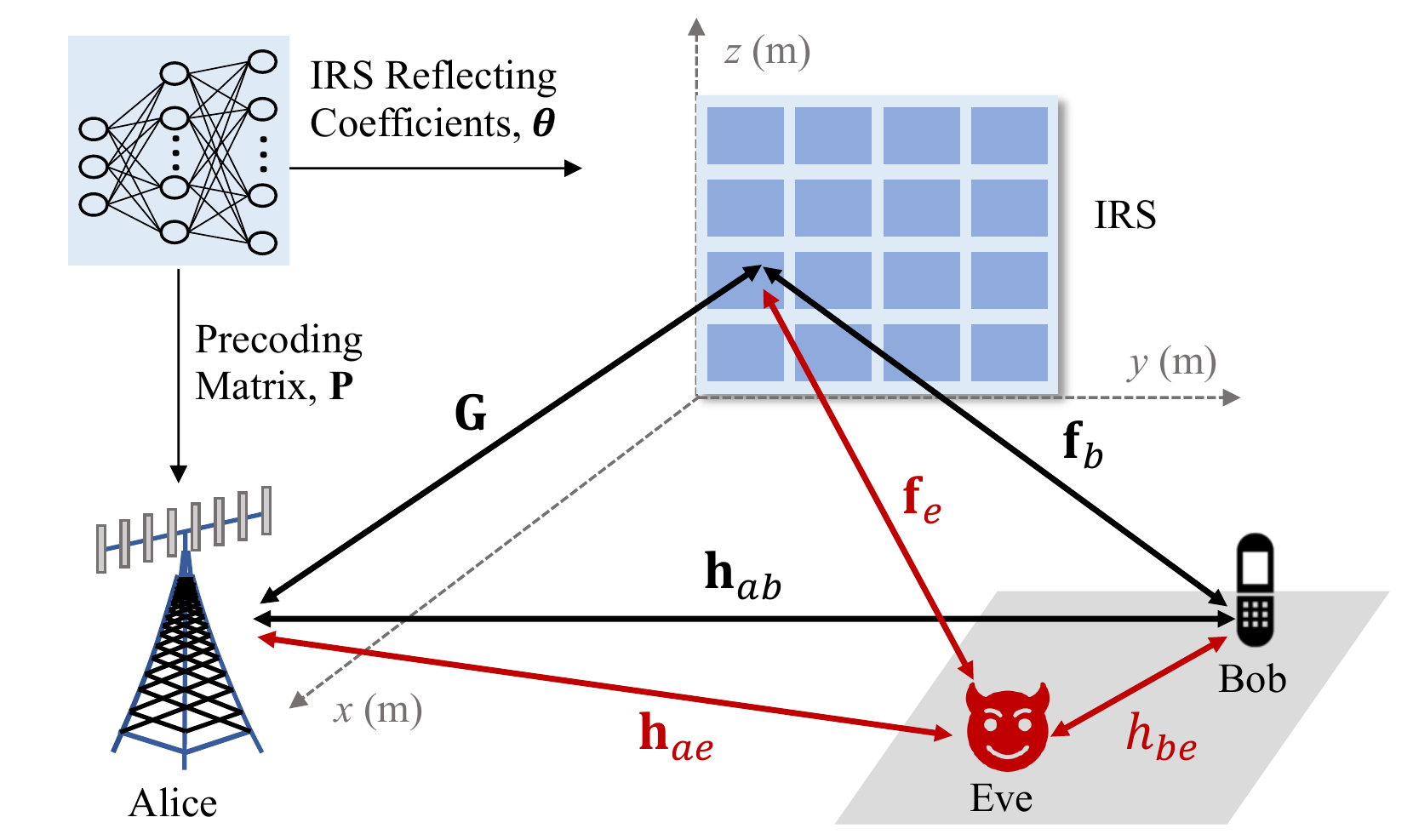}}
\caption{System model.}
\label{fig:system}
\end{figure}

We consider an IRS-assisted multi-antenna system as shown in Fig. \ref{fig:system}, which contains a multi-antenna BS, Alice, a single-antenna user equipment (UE), Bob, a single-antenna eavesdropper, Eve, and an IRS. We assume that the BS is a uniform linear array with $M$ antennas, and the IRS is a uniform planar array, which consists of $L=L_{\mathrm{H}}\times L_{\mathrm{V}}$ passive reflecting elements with $L_{\mathrm{H}}$ elements per row and $L_{\mathrm{V}}$ elements per column. To secure the communication, Alice and Bob perform PKG with the assistance of the IRS.

\begin{figure}[!t]
\centerline{\includegraphics[width=3.4in]{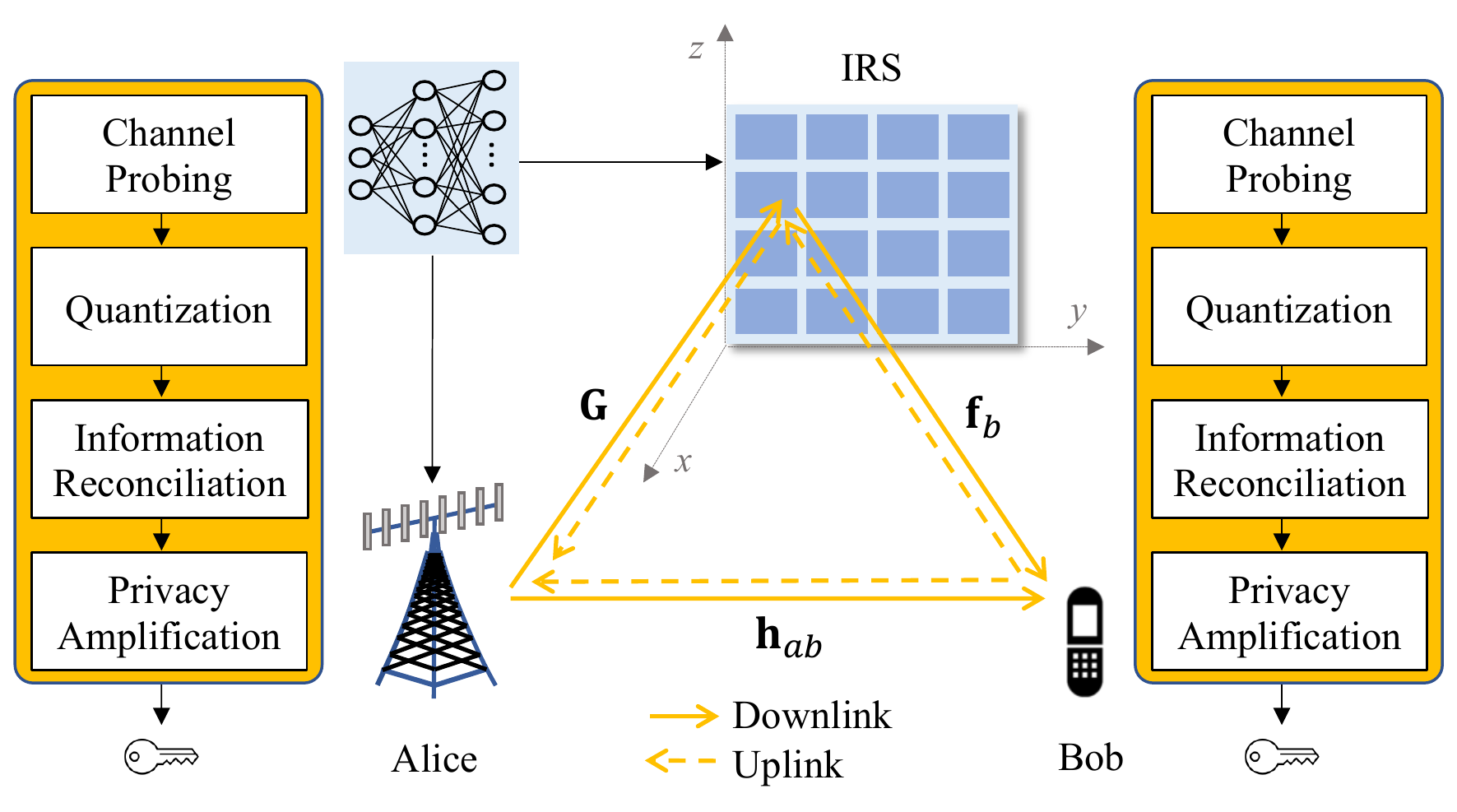}}
\caption{PKG protocol.}
\label{fig:channel}
\end{figure}

As shown in Fig. \ref{fig:channel}, the PKG protocol consists of four phases: channel probing, quantization, information reconciliation and privacy amplification. During the channel probing phase, Alice and Bob perform bidirectional channel measurements by transmitting pilot signals to each other.   We assume a time-division duplexing (TDD) protocol, and thus Alice and Bob can observe reciprocal channel information. During the quantization phase, Alice and Bob quantize the channel observations into binary key bits. Then the disagreed key bits are corrected by information reconciliation and the leaked information is eliminated by privacy amplification~\cite{Access1}. In this paper, we focus on the channel probing phase, because the channel measurements with IRS and multi-antenna configuration need to be optimized.

\subsection{Channel Model}
We denote the Alice-IRS, IRS-Bob (IRS-Eve), Alice-Bob (Alice-Eve) and Bob-Eve channels by $\mathbf{G}\in\mathbb{C}^{M\times L}$, $\mathbf{f}_{b}~(\mathbf{f}_{e})
 \in\mathbb{C}^{L\times 1}$, $\mathbf{h}_{ab}~(\mathbf{h}_{ae})\in\mathbb{C}^{M\times 1}$ and $h_{be}\in\mathbb{C}^{1\times 1}$, respectively. All the channels are modelled as spatially correlated Rician fading channels. To be specific, we have
\begin{align}
\mathbf{G} &= \mathbf{G}^{\mathrm{L}} + \mathbf{G}^{\mathrm{NL}}, \\
\mathbf{f}_{i}&= \mathbf{f}_{i}^{\mathrm{L}} + \mathbf{f}_{i}^{\mathrm{NL}}, i\in \{b,e\}, \\
\mathbf{h}_{ai}&= \mathbf{h}_{ai}^{\mathrm{L}} + \mathbf{h}_{ai}^{\mathrm{NL}}, i\in \{b,e\},  \\
h_{be} &= h_{be}^{\mathrm{L}} + h_{be}^{\mathrm{NL}},
\end{align}
where $\mathbf{G}^{\mathrm{L}}$, $\mathbf{f}_{i}^{\mathrm{L}}$, $\mathbf{h}_{ai}^{\mathrm{L}}$ and $h_{be}^{\mathrm{L}}$ are the line-of-sight (LoS) components of ${\mathbf{G}}$, ${\mathbf{f}_{i}}$, ${\mathbf{h}}_{ai}$ and $h_{be}$, respectively, and $\mathbf{G}^{\mathrm{NL}}$, $\mathbf{f}_{i}^{\mathrm{NL}}$, $\mathbf{h}_{ai}^{\mathrm{NL}}$ and $h_{be}^{\mathrm{NL}}$ are the non-line-of-sight (NLoS) components of ${\mathbf{G}}$, ${\mathbf{f}_{i}}$, ${\mathbf{h}}_{ai}$ and $h_{be}$, respectively. The LoS components are deterministic channels that depend on the locations of Alice, Bob and Eve, while the NLoS components are modelled as spatially correlated Rayleigh fading
channels. The NLoS components are further defined as 
\begin{align}
\mathbf{G}^{\mathrm{NL}}&=\mathbf{R}_{\mathrm{B}}^{\frac{1}{2}}\widetilde{\mathbf{G}}\mathbf{R}_{\mathrm{I}}^{\frac{1}{2}},  \\
\mathbf{f}_{i}^{\mathrm{NL}}&=\mathbf{R}_{\mathrm{I}}^{\frac{1}{2}}\widetilde{\mathbf{f}}_{i}, i\in \{b,e\}, \\
\mathbf{h}_{ai}^{\mathrm{NL}}&=\mathbf{R}_{\mathrm{B}}^{\frac{1}{2}}\widetilde{\mathbf{h}}_{ai}, i\in \{b,e\},
\end{align}
where $\mathbf{R}_{\mathrm{B}}\in\mathbb{R}^{M\times M}$ and $\mathbf{R}_{\mathrm{I}}\in\mathbb{R}^{L\times L}$ are the spatial correlation matrices at the BS and IRS, respectively, $\widetilde{\mathbf{G}}\in\mathbb{C}^{M\times L}$, $\widetilde{\mathbf{f}}_{i}\in\mathbb{C}^{L\times 1}$ and $\widetilde{\mathbf{h}}_{ai}\in\mathbb{C}^{M\times 1}$ have i.i.d. entries $\mathcal{CN}\left(0, \frac{\beta_{\mathrm{G}}}{1+\kappa_{\mathrm{G}}}\right)$, $\mathcal{CN}\left(0,\frac{\beta_{\mathrm{f}_{i}}}{1+\kappa_{\mathrm{f}_{i}}}\right)$ and $\mathcal{CN}\left(0, \frac{\beta_{\mathrm{h}_{ai}}}{1+\kappa_{\mathrm{h}_{ai}}}\right)$, respectively,
in which $\beta_{\mathrm{G}}$, $\beta_{\mathrm{f}_{i}}$ and $\beta_{\mathrm{h}_{ai}}$ are the path loss of ${\mathbf{G}}$, ${\mathbf{f}_{i}}$ and ${\mathbf{h}}_{ai}$, respectively, and $\kappa_{\mathrm{G}}$, $\kappa_{\mathrm{f}_{i}}$ and $\kappa_{\mathrm{h}_{ai}}$ are the Rician factors of ${\mathbf{G}}$, ${\mathbf{f}_{i}}$ and ${\mathbf{h}}_{ai}$, respectively. Moreover, $h_{be}^{\mathrm{NL}}\sim \mathcal{CN}\left(0, \frac{\beta_{\mathrm{h}_{be}}}{1+\kappa_{\mathrm{h}_{be}}}\right)$, where $\beta_{\mathrm{h}_{be}}$ and $\kappa_{\mathrm{h}_{be}}$ are the path loss and Rician factor of $\mathrm{h}_{be}$, respectively. Without loss of generality, we assume that the direct channels are in NLOS conditions, i.e., $\kappa_{\mathrm{h}_{ai}}=0$, and that $\kappa_{\mathrm{G}}=\kappa_{\mathrm{f}_{i}}=\kappa$.


 In practice, there exist spatial correlations among BS antennas and IRS elements. For simplicity, we adopt the spatial correlation model in isotropic scattering environments, where the ($n,m$)-th element of the spatial correlation matrix at the IRS is given by~\cite{WCL1}
\begin{align}
[\mathbf{R}_{\mathrm{I}}]_{n,m}=\frac{\sin(\frac{2\pi}{\lambda}||\mathbf{u}_n-\mathbf{u}_m||_2)}{\frac{2\pi}{\lambda}||\mathbf{u}_n-\mathbf{u}_m||_2},~n,m=1,\dots,L,
\end{align} 
where $\lambda$ is the wavelength, and $\mathbf{u}_n$ denotes the location of the $n$-th element and is computed as
$\mathbf{u}_n = \left[0, y_{n}\Delta, z_{n}\Delta\right]^{T}$, in which $y_{n}=\mod(n-1, L_{\mathrm{H}})$ and $z_{n}=\left \lfloor (n-1)/L_{\mathrm{H}} \right \rfloor$ are the horizontal and vertical indices of the $n$-th element, respectively, and $\Delta$ is the element spacing. The spatial correlation matrix at the BS is represented by $\mathbf{R}_{\mathrm{B}}$ with the ($n,m$)-th element given by 
\begin{align}
	[\mathbf{R}_{\mathrm{B}}]_{m,n}=\eta^{|m-n|},
\end{align}
 where $\eta$ is the correlation coefficient among BS antennas~\cite{GLOBECOM1}. 

\subsection{IRS Assisted Channel Probing}
 The channel probing is composed of bidirectional measurements, i.e., uplink and downlink probings. We assume a passive Eve who intends to eavesdrop on the information transmitted through legitimate channels. 
\subsubsection{Uplink Channel Probing}
During the uplink phase, Bob transmits a pilot signal $s_{u}\in \mathbb{C}^{1\times 1}$  to Alice. The received signals at Alice and Eve are given by
\begin{align} 
\mathbf{y}_{a} &= \sqrt{P_{b}}\left(\mathbf{h}_{ab} + \mathbf{G}\mathbf{\Theta}\mathbf{f}_{b}\right)s_{u} +\mathbf{n}_a,  \\
y_{be} &= \sqrt{P_{b}}\left(h_{be} + \mathbf{f}_{e}^{T}\mathbf{\Theta}\mathbf{f}_{b}\right)s_{u} + n_{e}^{u},
\end{align}
respectively, where $\mathbf{y}_a\in \mathbb{C}^{M\times 1}$, $P_{b}$ is the transmit power of Bob, $\mathbf{\Theta}=\text{diag}\left(e^{j\theta_{1}}, e^{j\theta_{2}}, \dots ,e^{j\theta_{L}}\right)$ denotes the IRS
 phase shift matrix with $\theta_{l}\in [0, 2\pi)$ for the $l$-th element, $\mathbf{n}_a\in \mathbb{C}^{M\times 1}$ is the complex Gaussian noise vector at Alice and $n_{e}^{u}$ is the complex Gaussian noise at Eve. Then Alice applies the precoding matrix $\mathbf{P}\in \mathbb{C}^{M\times N}$, where $N$ is the pilot length, to $\mathbf{y}_a$ and obtains
\begin{align} 
\bar{\mathbf{y}}_a =\sqrt{P_{b}}\mathbf{P}^{T} \left(\mathbf{h}_{ab} + \mathbf{G}\mathbf{\Theta}\mathbf{f}_{b}\right)s_{u} + \mathbf{P}^{T}\mathbf{n}_a.
\end{align}
After the least square (LS) channel estimation, the uplink channels at Alice and Eve are estimated as  
\begin{align} 
\widehat{\mathbf{y}}_a &=\sqrt{P_{b}}\mathbf{P}^{T} \left(\mathbf{h}_{ab} + \mathbf{G}\mathbf{\Theta}\mathbf{f}_{b}\right) + \mathbf{P}^{T}\mathbf{n}_a s_{u}^{*}, \\
\widehat{y}_{be} &= \sqrt{P_{b}}\left(h_{be} + \mathbf{f}_{e}^{T}\mathbf{\Theta}\mathbf{f}_{b}\right) +n_{e}^{u}s_{u}^{*}.
\end{align}
\subsubsection{Downlink Channel Probing}
During the downlink phase, Alice transmits a pilot signal of length $N$, $\mathbf{S}_{d}\in \mathbb{C}^{N\times N}$, with $\mathbf{S}_{d}^H\mathbf{S}_{d} = \mathbf{I}_{N}$ to Bob. The received signals at Bob and Eve are given by
\begin{align} 
\mathbf{y}_b &= \left(\mathbf{h}_{ab} + \mathbf{G}\mathbf{\Theta}\mathbf{f}_{b}\right)^{T}\mathbf{P}\mathbf{S}_{d}^{H} +\mathbf{n}_b, \\
\mathbf{y}_{ae} &= \left(\mathbf{h}_{ae} + \mathbf{G}\mathbf{\Theta}\mathbf{f}_{e}\right)^{T}\mathbf{P}\mathbf{S}_{d}^{H} +\mathbf{n}_{e}^{d},
\end{align}
respectively, where $\mathbf{y}_b, \mathbf{y}_{ae}\in \mathbb{C}^{1\times N}$,  $\mathbf{n}_b\in \mathbb{C}^{1\times N}$ is the complex Gaussian noise vector at Bob and $\mathbf{n}_{e}^{d}\in \mathbb{C}^{1\times N}$ is the complex Gaussian noise vector at Eve. After LS channel estimation, the downlink channels at Bob and Eve are estimated as  \begin{align} 
\widehat{\mathbf{y}}_b &= {\mathbf{y}}_b \mathbf{S}_{d}\left(\mathbf{S}_{d}^{H} \mathbf{S}_{d}\right)^{-1} = \left(\mathbf{h}_{ab} + \mathbf{G}\mathbf{\Theta}\mathbf{f}_{b}\right)^{T}\mathbf{P} + \mathbf{n}_b\mathbf{S}_{d}, \\
\widehat{\mathbf{y}}_{ae} &= {\mathbf{y}}_{ae}\mathbf{S}_{d}\left(\mathbf{S}_{d}^{H} \mathbf{S}_{d}\right)^{-1} = \left(\mathbf{h}_{ae} + \mathbf{G}\mathbf{\Theta}\mathbf{f}_{e}\right)^{T}\mathbf{P} +\mathbf{n}_{e}^{d}\mathbf{S}_{d}.
\end{align}
Then Bob and Eve transpose $\widehat{\mathbf{y}}_b$ and $\widehat{\mathbf{y}}_{ae}$, respectively, and obtain 
\begin{align}  
\widehat{\mathbf{y}}_b^{T} &= \mathbf{P}^{T}\left(\mathbf{h}_{ab} + \mathbf{G}\mathbf{\Theta}\mathbf{f}_{b}\right) + \mathbf{S}_{d}^{T}\mathbf{n}_b^{T}, \\
\widehat{\mathbf{y}}_{ae}^{T} &= \mathbf{P}^{T}\left(\mathbf{h}_{ae} + \mathbf{G}\mathbf{\Theta}\mathbf{f}_{e}\right) +\mathbf{S}_{d}^{T}\left(\mathbf{n}_{e}^{d}\right)^{T}.
\end{align}
To fully extract the randomness from BS antennas, we set $N=M$ in this paper.

PKG relies on the randomness of channels, and thus the deterministic LOS components do not contribute to the SKR~\cite{TIFS3}.
After estimating and removing the LOS components, the uplink channel observations are expressed as
\begin{align} \label{y_{a}}
\widetilde{\mathbf{y}}_a &=\sqrt{P_{b}}\mathbf{P}^{T} \left(\mathbf{h}_{ab}^{\mathrm{NL}} + \mathbf{G}^{\mathrm{NL}}\mathbf{\Theta}\mathbf{f}_{b}^{\mathrm{NL}}\right) + \mathbf{P}^{T}\mathbf{n}_a s_{u}^{*}, \\
\widetilde{y}_{be} &= \sqrt{P_{b}}\left(h_{be}^{\mathrm{NL}} + \left(\mathbf{f}_{e}^{\mathrm{NL}}\right)^{T}\mathbf{\Theta}\mathbf{f}_{b}^{\mathrm{NL}}\right) +n_{e}^{u}s_{u}^{*},
\end{align}
and the downlink channel observations are expressed as
\begin{align}  \label{y_{b}}
\widetilde{\mathbf{y}}_b^{T} &= \mathbf{P}^{T}\left(\mathbf{h}_{ab}^{\mathrm{NL}} + \mathbf{G}^{\mathrm{NL}}\mathbf{\Theta}\mathbf{f}_{b}^{\mathrm{NL}}\right) + \mathbf{S}_{d}^{T}\mathbf{n}_b^{T}, \\
\widetilde{\mathbf{y}}_{ae}^{T} &= \mathbf{P}^{T}\left(\mathbf{h}_{ae}^{\mathrm{NL}} + \mathbf{G}^{\mathrm{NL}}\mathbf{\Theta}\mathbf{f}_{e}^{\mathrm{NL}}\right) +\mathbf{S}_{d}^{T}\left(\mathbf{n}_{e}^{d}\right)^{T}. \label{y_{e}}
\end{align}

\section{Secret Key Rate}
\label{sec:secret_key rate}
SKR is defined as the maximum number of  secret key bits that can be extracted from a channel observation. The exact expression of SKR given the correlated eavesdropping channel is still an open problem. In this section, we derive a closed-form expression for the lower bound of SKR in the IRS-assisted multi-antenna system. 
 According to \cite{TIT1}, the lower bound of SKR is expressed as follows:
\begin{align} \label{SKR}
R_{\mathrm{sk}}\left(\widetilde{\mathbf{y}}_a,\widetilde{\mathbf{y}}_b, \widetilde{\mathbf{y}}_{ae}, \widetilde{y}_{be} \right)
\!=\! \mathrm{max}\big\{&I\left(\widetilde{\mathbf{y}}_a;\widetilde{\mathbf{y}}_b^{T}\right)\!-\! I\left(\widetilde{\mathbf{y}}_a;\widetilde{\mathbf{y}}_{ae}^{T}, \widetilde{y}_{be}\right), \nonumber \\
&I\left(\widetilde{\mathbf{y}}_a;\widetilde{\mathbf{y}}_b^{T}\right)\!-\! I\left(\widetilde{\mathbf{y}}_b;\widetilde{\mathbf{y}}_{ae}^{T}, \widetilde{y}_{be}\right)\big\},
\end{align}
where $I(X;Y)$ denotes the mutual information between random variables $X$ and $Y$. We assume that Eve gets as close as possible to Bob to maximize the correlation between his channel observation $\widetilde{\mathbf{y}}_{ae}^T$ and the legitimate channels. As such, $\widetilde{y}_{be}$ can be considered uncorrelated with the legitimate channels. For clarity, we denote  $\widetilde{\mathbf{y}}_{ae}$ by $\widetilde{\mathbf{y}}_{e}$ hereinafter. The lower bound of SKR in (\ref{SKR}) can therefore be simplified as 
\begin{align} \label{SKR_2}
R_{\mathrm{sk}}\left(\widetilde{\mathbf{y}}_a,\widetilde{\mathbf{y}}_b, \widetilde{\mathbf{y}}_{e} \right) = \mathrm{max}\big\{&I\left(\widetilde{\mathbf{y}}_a;\widetilde{\mathbf{y}}_b^{T}\right)- I\left(\widetilde{\mathbf{y}}_a;\widetilde{\mathbf{y}}_{e}^{T}\right), \nonumber \\
&I\left(\widetilde{\mathbf{y}}_a;\widetilde{\mathbf{y}}_b^{T}\right)- I\left(\widetilde{\mathbf{y}}_b;\widetilde{\mathbf{y}}_{e}^{T}\right)\big\}.
\end{align}

To facilitate further derivations, we introduce the cascaded channel $\mathbf{h}^{ai}_{\mathrm{c}}\in \mathbb{C}^{M(L+1)\times 1}$,  which is given by
\begin{align}
   \mathbf{h}^{ai}_{\mathrm{c}}=\text{vec}([\mathbf{h}_{ai}^{\mathrm{NL}}~\mathbf{G}^{\mathrm{NL}}\text{diag}(\mathbf{f}_{i}^{\mathrm{NL}})]), i\in \{b, e\}.
\end{align}
Then the combined channel can be expressed in a compact form:
\begin{align} \label{compact}
   \mathbf{h}_{ai}^{\mathrm{NL}} + \mathbf{G}^{\mathrm{NL}}\mathbf{\Theta}\mathbf{f}_{i}^{\mathrm{NL}}&\overset{(a)}{=}\begin{bmatrix}\mathbf{h}_{ai}^{\mathrm{NL}}~ \mathbf{G}^{\mathrm{NL}}\text{diag}(\mathbf{f}_{i}^{\mathrm{NL}})\end{bmatrix}
       \begin{bmatrix}
         1 \\
        \bm{\theta}
       \end{bmatrix} \nonumber \\
       &\overset{(b)}{=}(\widetilde{\bm{\theta}}^{T}\otimes\mathbf{I}_{M})\mathbf{h}^{ai}_{\mathrm{c}}, i\in \{b, e\},
\end{align}
where $\bm{\theta}=\left[[\mathbf{\Theta}]_{1,1}, [\mathbf{\Theta}]_{2,2},\cdots, [\mathbf{\Theta}]_{L,L}\right]^T$, $\widetilde{\bm{\theta}}=[1,\bm{\theta}^T]^T$, $(a)$ holds because $\text{diag}(\mathbf{\bm{\theta}})\mathbf{f}_{i}=\text{diag}(\mathbf{f}_{i})\bm{\theta}$ and $(b)$ holds because $\text{vec}(\mathbf{X}\mathbf{Y}\mathbf{Z})=(\mathbf{Z}^T\otimes \mathbf{X})\text{vec}(\mathbf{Y})$. Plugging (\ref{compact}) into (\ref{y_{a}}), (\ref{y_{b}}) and (\ref{y_{e}}), we have 
\begin{align} 
\widetilde{\mathbf{y}}_a &= \sqrt{P_{b}}\left(\widetilde{\bm{\theta}}^{T}\otimes\mathbf{P}^{T}\right)\mathbf{h}^{ab}_{\mathrm{c}} + \mathbf{P}^{T}\mathbf{n}_a s_{u}^{*} \nonumber \\
&= \sqrt{P_{b}}\left(\widetilde{\bm{\theta}}\otimes\mathbf{P}\right)^T\mathbf{h}^{ab}_{\mathrm{c}} + \mathbf{P}^{T}\mathbf{n}_a s_{u}^{*}, \\
\widetilde{\mathbf{y}}_b^{T} &= \left(\widetilde{\bm{\theta}}\otimes\mathbf{P}\right)^T\mathbf{h}^{ab}_{\mathrm{c}} + \mathbf{S}_{d}^{T}\mathbf{n}_b^{T}, \\
\widetilde{\mathbf{y}}_e^{T} &= \left(\widetilde{\bm{\theta}}\otimes\mathbf{P}\right)^T\mathbf{h}^{ae}_{\mathrm{c}} + \mathbf{S}_{d}^{T}\left(\mathbf{n}_{e}^{d}\right)^{T}. 
\end{align}

\begin{theorem} \label{theorem1}
The SKR of the IRS-assisted multi-antenna system is lower-bounded by 
\begin{align} \label{SKR_3}
R_{\mathrm{sk}}\left(\widetilde{\mathbf{y}}_a,\widetilde{\mathbf{y}}_b, \widetilde{\mathbf{y}}_{e} \right) = \mathrm{max}\big\{R_{\mathrm{sk}}^{1}\left(\widetilde{\mathbf{y}}_a,\widetilde{\mathbf{y}}_b, \widetilde{\mathbf{y}}_{e} \right), R_{\mathrm{sk}}^{2}\left(\widetilde{\mathbf{y}}_a,\widetilde{\mathbf{y}}_b, \widetilde{\mathbf{y}}_{e} \right)\big\},
\end{align}
in which $R_{\mathrm{sk}}^{1}\left(\widetilde{\mathbf{y}}_a,\widetilde{\mathbf{y}}_b, \widetilde{\mathbf{y}}_{e} \right)$ and $R_{\mathrm{sk}}^{2}\left(\widetilde{\mathbf{y}}_a,\widetilde{\mathbf{y}}_b, \widetilde{\mathbf{y}}_{e} \right)$ are given in (\ref{eq:Rsk1})  and (\ref{eq:Rsk2}) shown at the top of the next page, where $\delta^{2}$ is the Gaussian noise power, ${\mathbf{R}_{\mathrm{Z}}^{\mathrm{U}}}=\left(\widetilde{\bm{\theta}}\otimes\mathbf{P}\right)^T\mathbf{R}_{\mathrm{c}}^{\mathrm{U}}\left(\widetilde{\bm{\theta}}\otimes\mathbf{P}\right)^*$, $\mathbf{R}_{\mathrm{Z}}^{\mathrm{E}}=\left(\widetilde{\bm{\theta}}\otimes\mathbf{P}\right)^T\mathbf{R}_{\mathrm{c}}^{\mathrm{E}}\left(\widetilde{\bm{\theta}}\otimes\mathbf{P}\right)^*$ and $\mathbf{R}_{\mathrm{Z}}^{\mathrm{U},\mathrm{E}}=\left(\widetilde{\bm{\theta}}\otimes\mathbf{P}\right)^T\mathbf{R}_{\mathrm{c}}^{\mathrm{U},\mathrm{E}}\left(\widetilde{\bm{\theta}}\otimes\mathbf{P}\right)^*$ with
\begin{figure*}[t]
\begin{align} \label{eq:Rsk1}
&R_{\mathrm{sk}}^{1}\left(\widetilde{\mathbf{y}}_a,\widetilde{\mathbf{y}}_b, \widetilde{\mathbf{y}}_{e} \right) = 
\log_2 \left(\frac{\left|\mathbf{R}_{\mathrm{Z}}^{\mathrm{U}} + \delta^{2}\mathbf{I}_{M}\right|\left|\left(P_{b}\mathbf{R}_{\mathrm{Z}}^{\mathrm{U}} + \delta^{2}\mathbf{P}^{T}\mathbf{P}^{*}\right)\left(\mathbf{R}_{\mathrm{Z}}^{\mathrm{E}} + \delta^{2}\mathbf{I}_{M}\right) - P_{b}\left(\mathbf{R}_{\mathrm{Z}}^{\mathrm{U,E}}\right)^{2}\right|}{\left|\mathbf{R}_{\mathrm{Z}}^{\mathrm{E}} + \delta^{2}\mathbf{I}_{M}\right|\left|P_{b}\delta^{2}\mathbf{R}_{\mathrm{Z}}^{\mathrm{U}} + \delta^{2}\mathbf{P}^{T}\mathbf{P}^{*}\left(\mathbf{R}_{\mathrm{Z}}^{\mathrm{U}} + \delta^{2}\mathbf{I}_{M}\right)\right|}\right), \\
\label{eq:Rsk2}
&R_{\mathrm{sk}}^{2}\left(\widetilde{\mathbf{y}}_a,\widetilde{\mathbf{y}}_b, \widetilde{\mathbf{y}}_{e} \right) = \log_2 \left(\frac{\left|P_{b}\mathbf{R}_{\mathrm{Z}}^{\mathrm{U}} + \delta^{2}\mathbf{P}^{T}\mathbf{P}^{*}\right|\left|\left(\mathbf{R}_{\mathrm{Z}}^{\mathrm{U}}+\delta^{2}\mathbf{I}_{M}\right)\left(\mathbf{R}_{\mathrm{Z}}^{\mathrm{E}}+\delta^{2}\mathbf{I}_{M}\right) - \left(\mathbf{R}_{\mathrm{Z}}^{\mathrm{U,E}}\right)^2\right|}{\left|\mathbf{R}_{\mathrm{Z}}^{\mathrm{E}}+\delta^{2}\mathbf{I}_{M}\right|\left|P_{b}\delta^{2}\mathbf{R}_{\mathrm{Z}}^{\mathrm{U}} + \delta^{2}\mathbf{P}^{T}\mathbf{P}^{*}\left(\mathbf{R}_{\mathrm{Z}}^{\mathrm{U}} + \delta^{2}\mathbf{I}_{M}\right)\right|}\right),
\end{align}
\hrulefill
\end{figure*}
\begin{align}
&{\mathbf{R}_{\mathrm{c}}^{\mathrm{U}}}= \begin{bmatrix} \beta_{\mathrm{h}_{ab}}\mathbf{R}_{\mathrm{B}} & \mathbf{0}^{T} \\
    \mathbf{0} & \frac{\beta_{\mathrm{G}}\beta_{\mathrm{f}_{b}}}{\left({1+\kappa}\right)\left({1+\kappa}\right)}\mathbf{R}_{\mathrm{I}}\odot \mathbf{R}_{\mathrm{I}}\otimes \mathbf{R}_{\mathrm{B}}
  \end{bmatrix},        \\ &{\mathbf{R}_{\mathrm{c}}^{\mathrm{E}}}= \begin{bmatrix} \beta_{\mathrm{h}_{ae}}\mathbf{R}_{\mathrm{B}} & \mathbf{0}^{T} \\
    \mathbf{0} & \frac{\beta_{\mathrm{G}}\beta_{\mathrm{f}_{e}}}{\left({1+\kappa}\right)\left({1+\kappa}\right)}\mathbf{R}_{\mathrm{I}}\odot \mathbf{R}_{\mathrm{I}}\otimes \mathbf{R}_{\mathrm{B}}
  \end{bmatrix},        \\
&{\mathbf{R}_{\mathrm{c}}^{\mathrm{U,E}}}= \nonumber \\
&\begin{bmatrix} \rho\sqrt{\beta_{\mathrm{h}_{ab}}\beta_{\mathrm{h}_{ae}}} \mathbf{R}_{\mathrm{B}} \!&\! \!\mathbf{0}^{T}\! \\
    \mathbf{0} \!&\! \!\frac{\rho\beta_{\mathrm{G}}}{1+\kappa}\sqrt{\frac{\beta_{\mathrm{f}_{b}}\beta_{\mathrm{f}_{e}}}{({1+\kappa})({1+\kappa})}}\mathbf{R}_{\mathrm{I}}\odot \mathbf{R}_{\mathrm{I}}\otimes\mathbf{R}_{\mathrm{B}}
  \end{bmatrix},
\end{align}
where $\rho$ is the channel correlation coefficient
between Bob and Eve. For the considered Rayleigh fading channels, $\rho=J_{0}\left(\frac{2\pi d}{\lambda}\right)$, where $J_{0}(\cdot)$ is the zeroth order Bessel function of the first kind, $\lambda$ is the wavelength and $d$ is the distance between Eve and Bob \cite{CL2}.
\end{theorem}  
\begin{IEEEproof}
See Appendix \ref{AppendixA}.
\end{IEEEproof}

\section{Problem Formulation}
\label{sec:problem_formulation}
In this paper, we aim to maximize the SKR by jointly optimizing the BS's precoding matrix of $\mathbf{P}$ and the IRS's phase shift vector of  $\bm{\theta}$. The channel statistics of legitimate channels are assumed to be known by the BS~\cite{TWC1}. In reality, it is difficult to obtain the channel statistics of Eve. We first formulate the optimization problem for the practical case where the channel statistics of Eve are unknown (Case 1). We then formulate another optimization problem in the presence of Eve's channel statistics (Case 2) for comparison.

\subsection{Case 1: The Channel Statistics of Eve is Unknown}
In the absence of Eve’s channel statistics, we aim to maximize the mutual information between the channel observations of Alice and Bob, which is expressed as
\begin{align}
I\left(\widetilde{\mathbf{y}}_a;\widetilde{\mathbf{y}}_b^{T}\right) = \log_2 \left(\frac{\left|P_{b}\mathbf{R}_{\mathrm{Z}}^{\mathrm{U}} + \delta^{2}\mathbf{P}^{T}\mathbf{P}^{*}\right|\left|\mathbf{R}_{\mathrm{Z}}^{\mathrm{U}} + \delta^{2}\mathbf{I}_{M}\right|}{\left| P_{b}\delta^{2}\mathbf{R}_{\mathrm{Z}}^{\mathrm{U}} + \delta^{2}\mathbf{P}^{T}\mathbf{P}^{*}\left(\mathbf{R}_{\mathrm{Z}}^{\mathrm{U}} + \delta^{2}\mathbf{I}_{M}\right)\right|}\right),
\end{align}
where $\mathbf{R}_{\mathrm{Z}}^{\mathrm{U}}$ has been given in Theorem \ref{theorem1}. 

 Accordingly, the optimization problem is formulated as
\begin{align}
(\mathcal{P}1)~\mathop{\max_{\mathbf{P}, \bm{\theta}}}~&I\left(\widetilde{\mathbf{y}}_a;\widetilde{\mathbf{y}}_b^{T}\right)     \\
      s.t.~          &\text{Tr}\left(\mathbf{P}\mathbf{P}^H\right) = P_{\text{max}}, \tag{\ref{P}{a}}  \\
      &   \left|\bm{\theta}_{l}\right| = 1, l=1,2,\cdots,L, \tag{\ref{P}{b}}
\end{align}
where $P_{\text{max}}$ is the total transmit power constraint of Alice. 

\subsection{Case 2: The Channel Statistics of Eve is Known}
With the channel statistics of Eve, Alice can calculate the SKR using the expression derived in (\ref{SKR_3}).  Hence, the optimization problem is formulated as
\begin{align}
(\mathcal{P}2)~\mathop{\max_{\mathbf{P}, \bm{\theta}}}~&R_{\mathrm{sk}}\left(\widetilde{\mathbf{y}}_a,\widetilde{\mathbf{y}}_b, \widetilde{\mathbf{y}}_{e} \right)  \label{P}   \\
      s.t.~          &\text{Tr}\left(\mathbf{P}\mathbf{P}^H\right) = P_{\text{max}}, \tag{\ref{P}{a}} \label{c1a} \\
      &   \left|\bm{\theta}_{l}\right| = 1, l=1,2,\cdots,L. \label{c1b} \tag{\ref{P}{b}}
\end{align}

\section{Baseline Solutions}
\label{sec:baseline}
In this section, we develop water-filling algorithm-based baseline solutions for both optimization problems formulated  in the last section. 

\subsection{Case 1: The Channel Statistics of Eve is Unknown}
\label{sec:baseline_case1}
The original optimization problem ($\mathcal{P}1$) is a high-dimensional non-convex problem, which is difficult to solve directly with conventional optimization methods. For ease of further simplification, we re-derive $I\left(\widetilde{\mathbf{y}}_a;\widetilde{\mathbf{y}}_b^{T}\right)$ in the following lemma.
\begin{lemma} \label{lemma1}
$I\left(\widetilde{\mathbf{y}}_a;\widetilde{\mathbf{y}}_b^{T}\right)$ can be rewritten as
\begin{align} \label{eq:Iab2}
&I\left(\widetilde{\mathbf{y}}_a;\widetilde{\mathbf{y}}_b^{T}\right)=  \nonumber \\
& \log_2 \left(\frac{\left|P_{b}\delta_{\mathrm{U}}^{2}\mathbf{P}^{T}\mathbf{R}_{\mathrm{B}}\mathbf{P}^{*} \!+\! \delta^{2}\mathbf{P}^{T}\mathbf{P}^{*}\right|\left|\delta_{\mathrm{U}}^{2}\mathbf{P}^{T}\mathbf{R}_{\mathrm{B}}\mathbf{P}^{*} \!+\! \delta^{2}\mathbf{I}_{M}\right|}{\left| P_{b}\delta^{2}\delta_{\mathrm{U}}^{2}\mathbf{P}^{T}\mathbf{R}_{\mathrm{B}}\mathbf{P}^{*} \!+\! \delta^{2}\mathbf{P}^{T}\mathbf{P}^{*}\left(\delta_{\mathrm{U}}^{2}\mathbf{P}^{T}\mathbf{R}_{\mathrm{B}}\mathbf{P}^{*} \!+\! \delta^{2}\mathbf{I}_{M}\right)\right|}\right),
\end{align}
where $\delta_{\mathrm{U}}^{2}=\beta_{\mathrm{h}_{ab}} + \frac{\beta_{\mathrm{G}}\beta_{\mathrm{f}_{b}}}{\left({1+\kappa}\right)\left({1+\kappa}\right)}\bm{\theta}^H(\mathbf{R}_{\mathrm{I}}\odot\mathbf{R}_{\mathrm{I}})\bm{\theta}$.
\end{lemma}
\begin{IEEEproof}
See Appendix \ref{AppendixB}.
\end{IEEEproof}

We further rewrite $\mathbf{P}$ as $\mathbf{P}=\sqrt{\frac{P_{\text{max}}}{M}}\mathbf{P}_{\mathrm{e}}$, where $\mathbf{P}_{\mathrm{e}}$ is a normalized matrix with $\text{Tr}\left(\mathbf{P}_{\mathrm{e}}\mathbf{P}_{\mathrm{e}}^{H}\right)=M$.  Defining $P_{a}=\frac{P_{\text{max}}}{M}$, $I\left(\widetilde{\mathbf{y}}_a;\widetilde{\mathbf{y}}_b^{T}\right)$ can be derived as
\begin{align} \label{SKR2}
&{I}\left(\widetilde{\mathbf{y}}_a;\widetilde{\mathbf{y}}_b^{T}\right)= \nonumber \\ 
&\log_2 \left(\frac{\left|P_{b}P_{a}\delta_{\mathrm{U}}^{2}\widehat{\mathbf{R}}_{\mathrm{Z}} + \delta^{2}P_{a}\mathbf{P_{\mathrm{e}}}^{T}\mathbf{P}_{\mathrm{e}}^{*}\right|\left|P_{a}\delta_{\mathrm{U}}^{2}\widehat{\mathbf{R}}_{\mathrm{Z}} + \delta^{2}\mathbf{I}_{M}\right|}{\left| \delta^{2}P_{b}P_{a}\delta_{\mathrm{U}}^{2}\widehat{\mathbf{R}}_{\mathrm{Z}} + \delta^{2}P_{a}\mathbf{P_{\mathrm{e}}}^{T}\mathbf{P}_{\mathrm{e}}^{*}\left(P_{a}\delta_{\mathrm{U}}^{2}\widehat{\mathbf{R}}_{\mathrm{Z}} + \delta^{2}\mathbf{I}_{M}\right)\right|}\right),
\end{align}
where $\widehat{\mathbf{R}}_{\mathrm{Z}}=\mathbf{P_{\mathrm{e}}}^{T}\mathbf{R}_{\mathrm{B}}\mathbf{P}_{\mathrm{e}}^{*}$. To facilitate further optimization, we approximate $\mathbf{P_{\mathrm{e}}}^{T}\mathbf{P_{\mathrm{e}}}^{*}$ as $\mathbf{I}_{M}$ in the noise term. Accordingly, ${I}\left(\widetilde{\mathbf{y}}_a;\widetilde{\mathbf{y}}_b^{T}\right)$ is approximated as 
\begin{align} \label{SKR3}
&\widehat{I}\left(\widetilde{\mathbf{y}}_a;\widetilde{\mathbf{y}}_b^{T}\right)= \nonumber \\
&\log_2 \left(\frac{\left|P_{b}P_{a}\delta_{\mathrm{U}}^{2}\widehat{\mathbf{R}}_{\mathrm{Z}} + \delta^{2}P_{a}\mathbf{I}_{M}\right|\left|P_{a}\delta_{\mathrm{U}}^{2}\widehat{\mathbf{R}}_{\mathrm{Z}} + \delta^{2}\mathbf{I}_{M}\right|}{\left| \delta^{2}P_{b}P_{a}\delta_{\mathrm{U}}^{2}\widehat{\mathbf{R}}_{\mathrm{Z}} + \delta^{2}P_{a}\left(P_{a}\delta_{\mathrm{U}}^{2}\widehat{\mathbf{R}}_{\mathrm{Z}} + \delta^{2}\mathbf{I}_{M}\right)\right|}\right).
\end{align}

After Cholesky factorization, we have
\begin{align} \label{Cholesky}
\widehat{\mathbf{R}}_{\mathrm{Z}}
=\mathbf{P}_{\mathrm{e}}^T \mathbf{R}_{\mathrm{B}}^{\frac{1}{2}}\mathbf{R}_{\mathrm{B}}^{\frac{H}{2}} \mathbf{P}_{\mathrm{e}}^* = \left(\mathbf{R}_{\mathrm{B}}^{\frac{H}{2}}\mathbf{P}_{\mathrm{e}}^*\right)^H\left(\mathbf{R}_{\mathrm{B}}^{\frac{H}{2}}\mathbf{P}_{\mathrm{e}}^*\right).
\end{align}
 Then we perform the following eigenvalue decomposition:
 \begin{align} \label{SVD}
\mathbf{R}_{\mathrm{B}}^{\frac{H}{2}}\mathbf{P}_{\mathrm{e}}^*=\mathbf{U}\mathbf{\Lambda}\mathbf{U}^H,
\end{align}
where  $\mathbf{\Lambda}=\text{diag}(p_1,p_2,\dots,p_M)$ with the eigenvalues sorted in descending order. 
Substituting (\ref{Cholesky}) and (\ref{SVD}) into (\ref{SKR3}), $\widehat{I}\left(\widetilde{\mathbf{y}}_a;\widetilde{\mathbf{y}}_b^{T}\right)$ can be rewritten as (\ref{SKR2-scalar}), which is shown at the top of the next page.
\begin{figure*}[t]
\begin{align}
\label{SKR2-scalar}
\widehat{I}\left(\widetilde{\mathbf{y}}_a;\widetilde{\mathbf{y}}_b^{T}\right)= 
 \log_2 \left(\frac{\left|P_{b}P_{a}\delta_{\mathrm{U}}^{2}\mathbf{U}\mathbf{\Lambda}^T\mathbf{\Lambda}\mathbf{U}^H \!+\! \delta^{2}P_{a}\mathbf{I}_{M}\right|\left|P_{a}\delta_{\mathrm{U}}^{2}\mathbf{U}\mathbf{\Lambda}^T\mathbf{\Lambda}\mathbf{U}^H \!+\! \delta^{2}\mathbf{I}_{M}\right|}{\left| \delta^{2}P_{b}P_{a}\delta_{\mathrm{U}}^{2}\mathbf{U}\mathbf{\Lambda}^T\mathbf{\Lambda}\mathbf{U}^H \!+\! \delta^{2}P_{a}\left(P_{a}\delta_{\mathrm{U}}^{2}\mathbf{U}\mathbf{\Lambda}^T\mathbf{\Lambda}\mathbf{U}^H \!+\! \delta^{2}\mathbf{I}_{M}\right)\right|}\right) &= \sum_{i=1}^{M}\log_2 \left(\frac{(P_{b}\delta_{\mathrm{U}}^{2}p_{i}^{2} \!+\! \delta^{2})\left(P_{a}\delta_{\mathrm{U}}^{2}p_{i}^{2} \!+\! \delta^{2}\right)}{\delta^{2}P_{b}\delta_{\mathrm{U}}^{2}p_{i}^{2} \!+\! \delta^{2}P_{a}\delta_{\mathrm{U}}^{2}p_{i}^{2} \!+\! \delta^{4}}\right), 
\end{align}
\hrulefill
\end{figure*}

\begin{corollary} \label{corollary1}
$\widehat{I}\left(\widetilde{\mathbf{y}}_a;\widetilde{\mathbf{y}}_b^{T}\right)$ is maximized when all the IRS reflecting elements have the same phase configuration, i.e., $\theta_{i}=\theta_{j}, \forall i\ne j$.
\end{corollary}
\begin{IEEEproof}
See Appendix \ref{AppendixC}.
\end{IEEEproof}
Following Corollary \ref{corollary1}, we set $\theta_{i}=\theta_{j}, \forall i\ne j$ and have $\delta_{\mathrm{U}}^{2}=\beta_{\mathrm{h}_{ab}} + \frac{\beta_{\mathrm{G}}\beta_{\mathrm{f}_{b}}}{\left({1+\kappa}\right)\left({1+\kappa}\right)}\sum_{i=1}^{L}\sum_{j=1}^{L}\left[\mathbf{R}_{\mathrm{I}}\odot\mathbf{R}_{\mathrm{I}}\right]_{i,j}$.
We further perform eigenvalue decomposition on $\mathbf{R}_{\mathrm{B}}$ and obtain 
\begin{align}
\mathbf{R}_{\mathrm{B}}=\mathbf{U}_{\mathrm{B}}\mathbf{\Lambda}_{\mathrm{B}}\mathbf{U}_{\mathrm{B}}^H,
\end{align}
 where $\mathbf{\Lambda}_{\mathrm{B}}=\text{diag}\left(p_{\mathrm{B},1},p_{\mathrm{B},2},\dots,p_{\mathrm{B},M}\right)$ with the eigenvalues sorted in descending order. The optimal $\mathbf{\Lambda}$ can be obtained by solving the following optimization problem:
\begin{align} 
(\mathcal{P}1.1)~&\mathop{\max_{p_{i}}}~\widehat{I}\left(\widetilde{\mathbf{y}}_a;\widetilde{\mathbf{y}}_b^{T}\right) \label{P2}  \\
      &s.t.~\sum_{i=1}^{M}\frac{p_{i}^{2}}{p_{\mathrm{B},i}}=M, \label{c2} \tag{\ref{P2}{a}}
\end{align}
where (\ref{c2}) comes from the constraint $\text{Tr}\left(\mathbf{P}_{\mathrm{e}}\mathbf{P}_{\mathrm{e}}^H\right) =\text{Tr}(\mathbf{\Lambda}_\mathrm{B}^{-1}\mathbf{\Lambda}^2)= M$. ($\mathcal{P}1.1$) can be solved by using water-filling algorithm~\cite{TSP1}. Denoting the optimal $\mathbf{\Lambda}$ obtained from ($\mathcal{P}1.1$) by $\mathbf{\Lambda}^{\mathrm{opt}}$, the corresponding $\mathbf{P}_{\mathrm{e}}$ can be computed from (\ref{SVD}) and expressed as $\mathbf{P}_{\mathrm{e}}^{\mathrm{opt}}= \left(\mathbf{R}_\mathrm{B}^{-\frac{1}{2}}\mathbf{U}\mathbf{\Lambda}^{\mathrm{opt}}\mathbf{U}^H\right)^{*}$. Note that $\mathbf{P}_{\mathrm{e}}^{\mathrm{opt}}$ is suboptimal to the original optimization problem  ($\mathcal{P}1$) due to the approximation in (\ref{SKR3}). 

\subsection{Case 2: The Channel Statistics of Eve is Known}
\label{sec:baseline_case2}
With the channel statistics of Eve, $R_{\mathrm{sk}}\left(\widetilde{\mathbf{y}}_a,\widetilde{\mathbf{y}}_b, \widetilde{\mathbf{y}}_{e} \right)$ is the maximum between $R_{\mathrm{sk}}^{1}\left(\widetilde{\mathbf{y}}_a,\widetilde{\mathbf{y}}_b, \widetilde{\mathbf{y}}_{e} \right)$ and $R_{\mathrm{sk}}^{2}\left(\widetilde{\mathbf{y}}_a,\widetilde{\mathbf{y}}_b, \widetilde{\mathbf{y}}_{e} \right)$. To maximize $R_{\mathrm{sk}}\left(\widetilde{\mathbf{y}}_a,\widetilde{\mathbf{y}}_b, \widetilde{\mathbf{y}}_{e} \right)$, we first obtain the optimal $R_{\mathrm{sk}}^{1}\left(\widetilde{\mathbf{y}}_a,\widetilde{\mathbf{y}}_b, \widetilde{\mathbf{y}}_{e} \right)$ and $R_{\mathrm{sk}}^{2}\left(\widetilde{\mathbf{y}}_a,\widetilde{\mathbf{y}}_b, \widetilde{\mathbf{y}}_{e} \right)$, denoted by $R_{\mathrm{sk}}^{1, \text{max}}\left(\widetilde{\mathbf{y}}_a,\widetilde{\mathbf{y}}_b, \widetilde{\mathbf{y}}_{e} \right)$ and $R_{\mathrm{sk}}^{2, \text{max}}\left(\widetilde{\mathbf{y}}_a,\widetilde{\mathbf{y}}_b, \widetilde{\mathbf{y}}_{e} \right)$, respectively. Then the optimal $R_{\mathrm{sk}}\left(\widetilde{\mathbf{y}}_a,\widetilde{\mathbf{y}}_b, \widetilde{\mathbf{y}}_{e} \right)$, denoted by $R_{\mathrm{sk}}^{\text{max}}\left(\widetilde{\mathbf{y}}_a,\widetilde{\mathbf{y}}_b, \widetilde{\mathbf{y}}_{e} \right)$, is the maximum between $R_{\mathrm{sk}}^{1,\text{max}}\left(\widetilde{\mathbf{y}}_a,\widetilde{\mathbf{y}}_b, \widetilde{\mathbf{y}}_{e} \right)$ and $R_{\mathrm{sk}}^{2,\text{max}}\left(\widetilde{\mathbf{y}}_a,\widetilde{\mathbf{y}}_b, \widetilde{\mathbf{y}}_{e} \right)$. Following similar steps in the last subsection, we approximate (\ref{eq:Rsk1}) and (\ref{eq:Rsk2}) as (\ref{eq:Rsk12}) and (\ref{eq:Rsk22}), respectively, 
\begin{figure*}[t]
\begin{align} \label{eq:Rsk12}
\widehat{R}_{\mathrm{sk}}^{1}\left(\widetilde{\mathbf{y}}_a,\widetilde{\mathbf{y}}_b, \widetilde{\mathbf{y}}_{e} \right) &= 
\log_2 \left(\frac{\left|P_{a}\delta_{\mathrm{U}}^{2}\widehat{\mathbf{R}}_{\mathrm{Z}} + \delta^{2}\mathbf{I}_{M}\right|\left|\left(P_{b}P_{a}\delta_{\mathrm{U}}^{2}\widehat{\mathbf{R}}_{\mathrm{Z}} + \delta^{2}P_{a}\mathbf{I}_{M}\right)\left(P_{a}\delta_{\mathrm{E}}^{2}\widehat{\mathbf{R}}_{\mathrm{Z}} + \delta^{2}\mathbf{I}_{M}\right) - P_{b}\left(P_{a}\delta_{\mathrm{U,E}}^{2}\widehat{\mathbf{R}}_{\mathrm{Z}}\right)^{2}\right|}{\left|P_{a}\delta_{\mathrm{E}}^{2}\widehat{\mathbf{R}}_{\mathrm{Z}} + \delta^{2}\mathbf{I}_{M}\right|\left|\delta^{2}P_{b}P_{a}\delta_{\mathrm{U}}^{2}\widehat{\mathbf{R}}_{\mathrm{Z}} + \delta^{2}P_{a}\left(P_{a}\delta_{\mathrm{U}}^{2}\widehat{\mathbf{R}}_{\mathrm{Z}} + \delta^{2}\mathbf{I}_{M}\right)\right|}\right) \nonumber \\
&= \sum_{i=1}^{M}\log_2 \left(\frac{\left(P_{a}\delta_{\mathrm{U}}^{2}p_{i}^{2}+ \delta^{2}\right)\left(P_{b}\delta_{\mathrm{U}}^{2}p_{i}^{2} + \delta^{2} - \frac{P_b P_a\delta_{\mathrm{U,E}}^{4}p_{i}^4}{P_a\delta_{\mathrm{E}}^{2}p_{i}^2 + \delta^{2}}\right)}{\delta^{2}P_{b}\delta_{\mathrm{U}}^{2}p_{i}^{2}+ \delta^{2}P_{a}\delta_{\mathrm{U}}^{2}p_{i}^{2} + \delta^{4}}\right),
\\
\label{eq:Rsk22}
\widehat{R}_{\mathrm{sk}}^{2}\left(\widetilde{\mathbf{y}}_a,\widetilde{\mathbf{y}}_b, \widetilde{\mathbf{y}}_{e} \right) &= \log_2 \left(\frac{|P_{b}P_{a}\delta_{\mathrm{U}}^{2}\widehat{\mathbf{R}}_{\mathrm{Z}} + \delta^{2}P_a\mathbf{I}_{M}|\left|\left(P_{a}\delta_{\mathrm{U}}^{2}\widehat{\mathbf{R}}_{\mathrm{Z}} + \delta^{2}\mathbf{I}_{M}\right)\left(P_{a}\delta_{\mathrm{E}}^{2}\widehat{\mathbf{R}}_{\mathrm{Z}}+\delta^{2}\mathbf{I}_{M}\right) - \left(P_{a}\delta_{\mathrm{U,E}}^{2}\widehat{\mathbf{R}}_{\mathrm{Z}}\right)^2\right|}{|P_{a}\delta_{\mathrm{E}}^{2}\widehat{\mathbf{R}}_{\mathrm{Z}} + \delta^{2}\mathbf{I}_{M}|\left|\delta^{2}P_{b}P_{a}\delta_{\mathrm{U}}^{2}\widehat{\mathbf{R}}_{\mathrm{Z}} + \delta^{2}P_{a}\left(P_{a}\delta_{\mathrm{U}}^{2}\widehat{\mathbf{R}}_{\mathrm{Z}} + \delta^{2}\mathbf{I}_{M}\right)\right|}\right) \nonumber \\
&=\sum_{i=1}^{M}\log_2 \left(\frac{(P_{b}\delta_{\mathrm{U}}^{2}p_{i}^{2} + \delta^{2})\left(P_{a}\delta_{\mathrm{U}}^{2}p_{i}^{2}+ \delta^{2} - \frac{P_a^2\delta_{\mathrm{U,E}}^{4}p_{i}^4}{P_a\delta_{\mathrm{E}}^{2}p_{i}^2 + \delta^{2}}\right)}{\delta^{2}P_{b}\delta_{\mathrm{U}}^{2}p_{i}^{2}+ \delta^{2}P_{a}\delta_{\mathrm{U}}^{2}p_{i}^{2} + \delta^{4}}\right),
\end{align}
\hrulefill
\end{figure*}
where $\delta_{\mathrm{E}}^{2}=\beta_{\mathrm{h}_{ae}}+\frac{\beta_{\mathrm{G}}\beta_{\mathrm{f}_{e}}}{\left({1+\kappa}\right)\left({1+\kappa}\right)}\sum_{i=1}^{L}\sum_{j=1}^{L}\left[\mathbf{R}_{\mathrm{I}}\odot\mathbf{R}_{\mathrm{I}}\right]_{i,j}$  and $\delta_{\mathrm{U,E}}^{2} = \rho\sqrt{\beta_{\mathrm{h}_{ab}}\beta_{\mathrm{h}_{ae}}} + \frac{\rho\beta_{\mathrm{G}}}{1 + \kappa}\sqrt{\frac{\beta_{\mathrm{f}_{b}}\beta_{\mathrm{f}_{e}}}{({1 + \kappa})({1 + \kappa})}}\sum_{i=1}^{L}\sum_{j=1}^{L}\left[\mathbf{R}_{\mathrm{I}}\odot\mathbf{R}_{\mathrm{I}}\right]_{i,j}$. Then we formulate the following two optimization problems:
\begin{align} 
(\mathcal{P}2.1)~&\mathop{\max_{p_{i}}}~\widehat{R}_{\mathrm{sk}}^{1}\left(\widetilde{\mathbf{y}}_a,\widetilde{\mathbf{y}}_b, \widetilde{\mathbf{y}}_{e} \right) \label{P2.1}  \\
      &s.t.~\sum_{i=1}^{M}\frac{p_{i}^{2}}{p_{\mathrm{B},i}}=M, \label{c2.1} \tag{\ref{P2.1}{a}} \\
(\mathcal{P}2.2)~&\mathop{\max_{p_{i}}}~\widehat{R}_{\mathrm{sk}}^{2}\left(\widetilde{\mathbf{y}}_a,\widetilde{\mathbf{y}}_b, \widetilde{\mathbf{y}}_{e} \right) \label{P2.2}  \\
      &s.t.~\sum_{i=1}^{M}\frac{p_{i}^{2}}{p_{\mathrm{B},i}}=M, \label{c2.2} \tag{\ref{P2.2}{a}}
\end{align}
which can be solved by using water-filling
algorithm. Let $\mathbf{\Lambda}^{\mathrm{opt},1}$ and $\mathbf{\Lambda}^{\mathrm{opt},2}$ denote the optimal $\mathbf{\Lambda}$ obtained from ($\mathcal{P}2.1$) and ($\mathcal{P}2.2$), respectively. Accordingly, the optimal $\mathbf{P}_{\mathrm{e}}$ for ($\mathcal{P}2.1$) and ($\mathcal{P}2.2$) is computed by 
$\mathbf{P}_{\mathrm{e}}^{\mathrm{opt},1}= \left(\mathbf{R}_\mathrm{B}^{-\frac{1}{2}}\mathbf{U}\mathbf{\Lambda}^{\mathrm{opt},1}\mathbf{U}^H\right)^{*}$ and $\mathbf{P}_{\mathrm{e}}^{\mathrm{opt},2}= \left(\mathbf{R}_\mathrm{B}^{-\frac{1}{2}}\mathbf{U}\mathbf{\Lambda}^{\mathrm{opt},2}\mathbf{U}^H\right)^{*}$, respectively. $R_{\mathrm{sk}}^{1, \text{max}}\left(\widetilde{\mathbf{y}}_a,\widetilde{\mathbf{y}}_b, \widetilde{\mathbf{y}}_{e} \right)$ and $R_{\mathrm{sk}}^{2, \text{max}}\left(\widetilde{\mathbf{y}}_a,\widetilde{\mathbf{y}}_b, \widetilde{\mathbf{y}}_{e} \right)$ are obtained by substituting $\mathbf{P}_{\mathrm{e}}^{\mathrm{opt},1}$ and $\mathbf{P}_{\mathrm{e}}^{\mathrm{opt},2}$ into (\ref{eq:Rsk1}) and (\ref{eq:Rsk2}), respectively. Finally, we have $R_{\mathrm{sk}}^{\text{max}}\left(\widetilde{\mathbf{y}}_a,\widetilde{\mathbf{y}}_b, \widetilde{\mathbf{y}}_{e} \right) = \text{max}\left\{R_{\mathrm{sk}}^{1,\text{max}}\left(\widetilde{\mathbf{y}}_a,\widetilde{\mathbf{y}}_b, \widetilde{\mathbf{y}}_{e} \right), R_{\mathrm{sk}}^{2,\text{max}}\left(\widetilde{\mathbf{y}}_a,\widetilde{\mathbf{y}}_b, \widetilde{\mathbf{y}}_{e} \right)\right\}$.

\section{Proposed Machine Learning-Based Method}
\label{sec:machine_learning}
The baseline solution is an iterative algorithm with a complex structure. In this section, we propose to directly solve ($\mathcal{P}1$) using DNNs. More specifically, we aim to train a DNN to output the optimal transmit precoding matrix, $\mathbf{P}$, and the phase shift vector at the IRS, $\bm{\theta}$, that maximize the SKR based on the location information, transmit power, and channel statics.

\subsection{Case 1: The Channel Statistics of Eve is Unknown}
\label{sec:PKG_Net_case1}
\begin{figure}[!t]
\centerline{\includegraphics[width=3.4in]{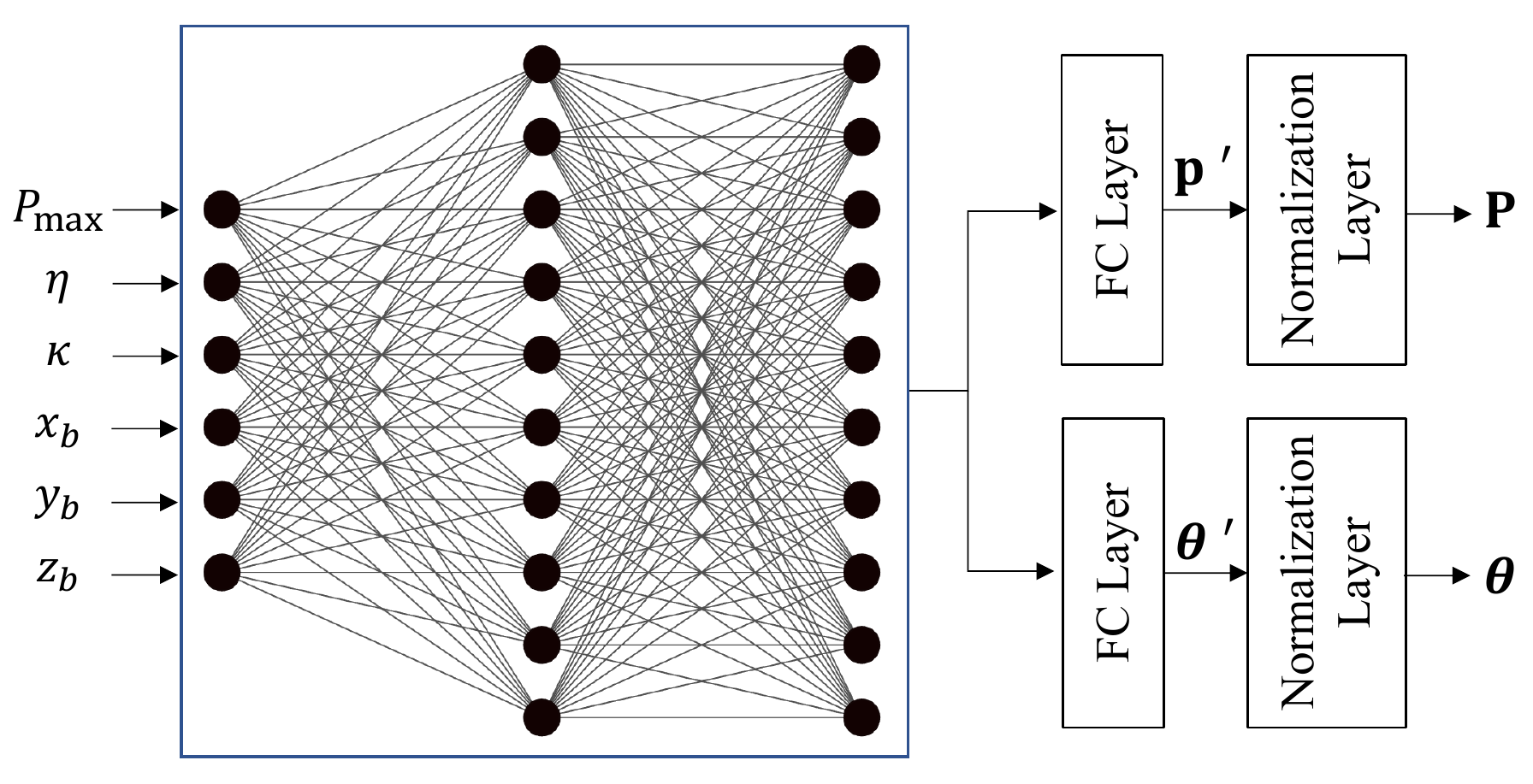}}
\caption{Structure of the proposed PKG-Net in the absence of Eve information.}
\label{fig:network}
\end{figure}

In the absence of Eve's channel statistics, the structure of the proposed PKG-Net is shown in Fig.~\ref{fig:network}. 
In particular, the input consists of the location information of Bob $\left(x_{b}, y_{b}, z_{b}\right)$, the maximum transmit power $P_{\mathrm{max}}$, the correlation coefficient among BS antennas $\eta$ and the  Rician factor $\kappa$,
which is a $6\times 1$ vector. Then two fully connected (FC) layers are adopted as hidden layers for feature extraction with ReLu as the activation function. The output layer consists of two FC layers. Since neural networks only support real-valued outputs, the first $2\times M\times M$ FC layer outputs a real-valued vector $\mathbf{p}^{'} \in \mathbb{R}^{2M^2\times 1}$ and the second $2\times L$ FC layer outputs a real-valued vector $\bm{\theta}^{'}\in \mathbb{R}^{2L\times 1}$. To obtain the precoding matrix and satisfy the total power constraint, the first normalization layer converts $\mathbf{p}^{'}$ to a complex-valued matrix $\mathbf{P}^{''}$ and performs the following normalization step: 
\begin{align}
\mathbf{P} = \sqrt{MP_{a}}\frac{\mathbf{P}^{''}}{\|\mathbf{P}^{''}\|_F^2}.
\end{align}
Similarly, to satisfy the unit modulus constraint, the second normalization layer performs 
\begin{align}
\bm{\theta}_{l} = \frac{\bm{\theta}^{'}_{l}}{\sqrt{\left(\bm{\theta}^{'}_{l}\right)^{2} + \left(\bm{\theta}^{'}_{l+L}\right)^{2}}} + j \frac{\bm{\theta}^{'}_{l+L}}{\sqrt{\left(\bm{\theta}^{'}_{l}\right)^{2} + \left(\bm{\theta}^{'}_{l+L}\right)^{2}}}, \forall l.
\end{align}

During the training phase, the PKG-Net learns to update its parameters in an unsupervised manner. The goal is to maximize $I\left(\widetilde{\mathbf{y}}_a;\widetilde{\mathbf{y}}_b^{T}\right)$, namely, minimize the following loss function: \begin{align}
\mathcal{L}_{\mathrm{loss}}^{1} = -\frac{1}{K}\sum_{k=1}^{K}I\left(\widetilde{\mathbf{y}}_{a,k};\widetilde{\mathbf{y}}_{b,k}^{T}\right),
\end{align}
where $K$ is the number of training samples, and $\widetilde{\mathbf{y}}_{a,k}$ and $\widetilde{\mathbf{y}}_{b,k}^{T}$ are the channel estimations in the $k$-th training. It is clear that the smaller the loss function, the higher the average SKR. {Recall that $\widetilde{\mathbf{y}}_{a}$ and $\widetilde{\mathbf{y}}_{b}^{T}$ are functions of $\mathbf{P}$ and $\mathbf{\theta}$, which are the outputs of PKG-Net. During the training, PKG-Net learns to obtain the optimal $\mathbf{P}$ and $\mathbf{\theta}$ that minimizes the loss function.} The training phase is performed offline, and thus the computational complexity is less of a concern.

\subsection{Case 2: The Channel Statistics of Eve is Known}
\begin{figure}[!t]
\centerline{\includegraphics[width=3.4in]{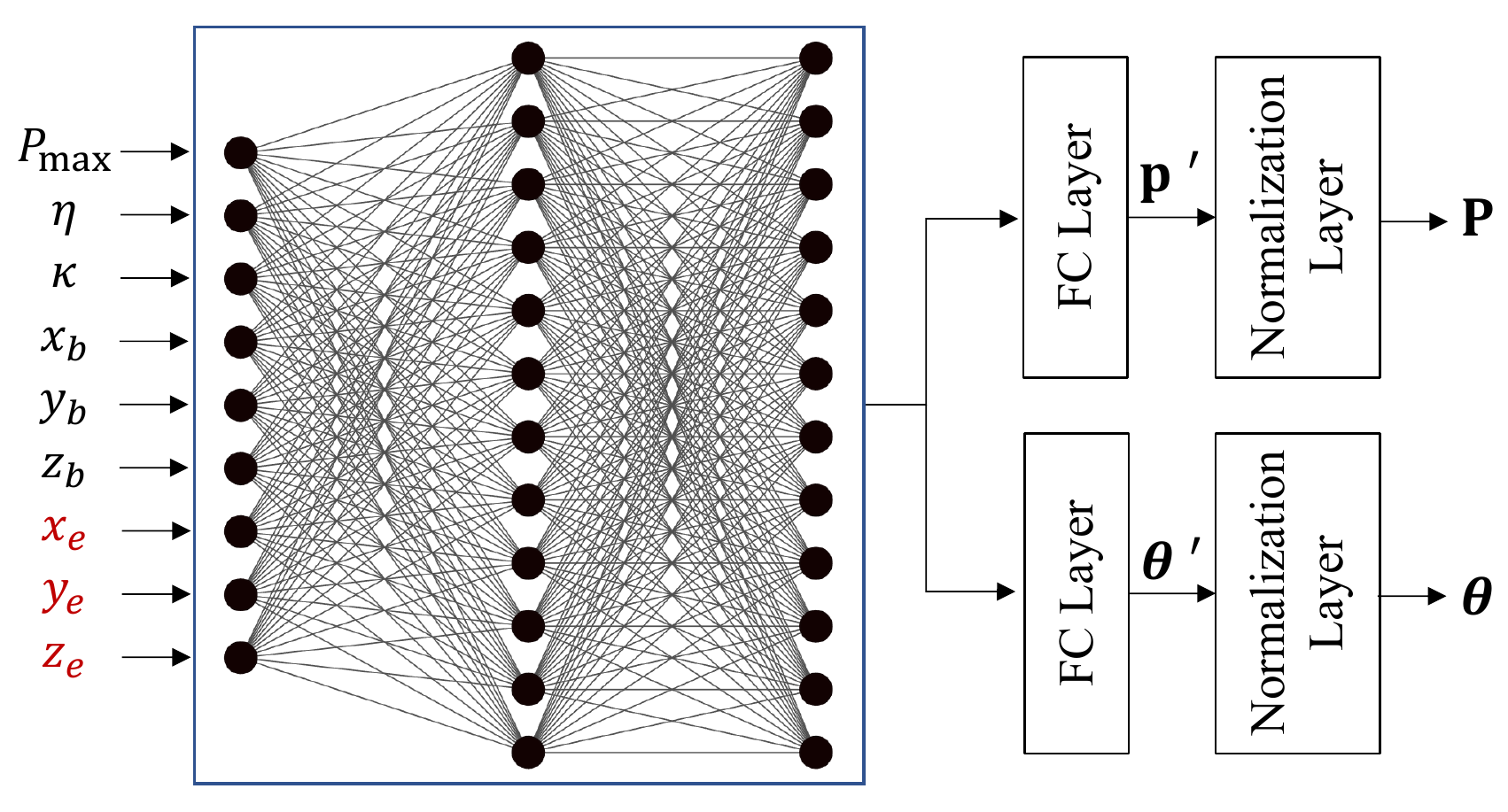}}
\caption{Structure of the proposed PKG-Net with Eve information. }
\label{fig:network_eve}
\end{figure}

In the presence of Eve's channel statistics, the structure of the proposed PKG-Net is shown in Fig.~\ref{fig:network_eve}. Besides the location of Bob, transmit power and channel statistics, we also treat the location information of Eve, i.e., $\left(x_{e}, y_{e}, z_{e}\right)$, as input.   The hidden layers and output layer have the same structure as the network presented in the previous subsection. During the training phase, the goal is to maximize $R_{\mathrm{sk}}\left(\widetilde{\mathbf{y}}_a,\widetilde{\mathbf{y}}_b, \widetilde{\mathbf{y}}_{e} \right)$, namely, minimize the following loss function:
\begin{align}
\mathcal{L}_{\mathrm{loss}}^{2} = -\frac{1}{K}\sum_{k=1}^{K}R_{\mathrm{sk}}\left(\widetilde{\mathbf{y}}_{a,k},\widetilde{\mathbf{y}}_{b,k}, \widetilde{\mathbf{y}}_{e,k} \right).
\end{align}

In both cases, during the online inference phase, the BS directly designs the precoding matrix and the IRS phase shifts according to the output of the trained neural network as soon as it gets the input information. 


\section{Simulation Results}
\label{sec:simulation}
In this section, we numerically evaluate the SKR performance of the proposed
PKG-Net in comparison with the benchmark methods.

\subsection{Simulation Setup}
The coordinates of the BS and the IRS in meters are set as (5, -30, 0) and (0, 0, 0), respectively. The large-scale path loss in dB is computed by 
\begin{align}
	\beta_{i}= \beta_{0,i} - 10\alpha_{i}\text{log}_{10}\left(d/d_{0}\right),
\end{align}
where $\beta_{0,i}$ is the path loss at the reference distance, $d_{0}=1$~m is the reference distance, $d$ is the transmission distance, and $\alpha_{i}$ is the path-loss exponent. 
For the direct link, i.e., $i=\mathrm{h}$, we set $\beta_{0,i}= -32.6$ dB and $\alpha_{i}=3.67$; for the reflecting links, i.e., $i=\mathrm{G}$ or $\mathrm{f}$, we set $\beta_{0,i}= -30$ dB and $\alpha_{i}=2.2$. The channel bandwidth is set to $W=20$ MHz and the noise power is computed by $\delta^{2}=-174\ \mathrm{dBm/Hz} + 10 \mathrm{log}10(W) + 10\ \mathrm{dB}$ \cite{buzzi2021ris, WCL1}. The IRS is assumed to be a uniform square array, i.e., $L_{\mathrm{H}}=L_{\mathrm{V}}$. Unless otherwise specified, the transmit power of Bob is set to $P_{b}=10$ dBm, the default spatial correlation coefficient at the BS is set to $\eta=0.4$ and the IRS element spacing is half a wavelength, i.e., $\Delta=\frac{\lambda}{2}$. 

During the training of the proposed
PKG-Net in both cases, the UE location is uniformly distributed in the $xy$-plane with $x\in [5, 15]$ and  $y\in [5, 15]$ to achieve generalization ability to various UE  locations. Moreover, $P_{\mathrm{max}}$ is uniformly distributed in $[10, 30]$ dBm, $\eta$ is uniformly distributed in $[0, 1]$, and $\kappa$ is uniformly distributed in $[0, 10]$. In particular, in case 2, the location of Eve is uniformly distributed in the $xy$-plane with $x\in [5, 15]$ and  $y\in [5, 15]$.
In both cases, each hidden layer of the PKG-Net has 100 neurons. In each training epoch, 1000 random UE locations are used as training samples with a batch size of 100. The DNN is implemented using TensorFlow and updated by Adam optimizer with a learning rate of 0.001. The training is terminated when the loss function on the verification data set does not {decrease} by more than 10 consecutive epochs or the number of training epochs reaches 100.

The proposed algorithms and benchmark methods are summarized as follows:
\begin{itemize}
\item \textbf{PKG-Net w/o Eve info}: Represent the PKG-Net in the absence of Eve's channel statistics proposed in Section \ref{sec:machine_learning}.

\item \textbf{PKG-Net w/ Eve info}: Represent the PKG-Net with Eve information proposed in Section \ref{sec:machine_learning}.

\item \textbf{Baseline solution w/o Eve info}: Represent the water-filling algorithm-based solution proposed in Section \ref{sec:baseline_case1}.

\item \textbf{Baseline solution w/ Eve info}: Represent the water-filling algorithm-based solution proposed in Section \ref{sec:baseline_case2}.

\item \textbf{Random configuration}: Represent a non-optimized scheme, where both the precoding matrix $\mathbf{P}$ and the IRS phase shift vector $\bm{\theta}$ are randomly configured.
\end{itemize}

\subsection{Results}
In the following numerical evaluation, the UE location is set to $\left(x_{b}, y_{b}, z_{b}\right)=(10, 10, 0)$ and the Eve location is set to $\left(x_{e}, y_{e}, z_{e}\right)=(x_{b}-0.5\lambda, y_{b}, z_{b})$ unless otherwise stated. 

\begin{figure}[!t]
\centerline{\includegraphics[width=3.4in]{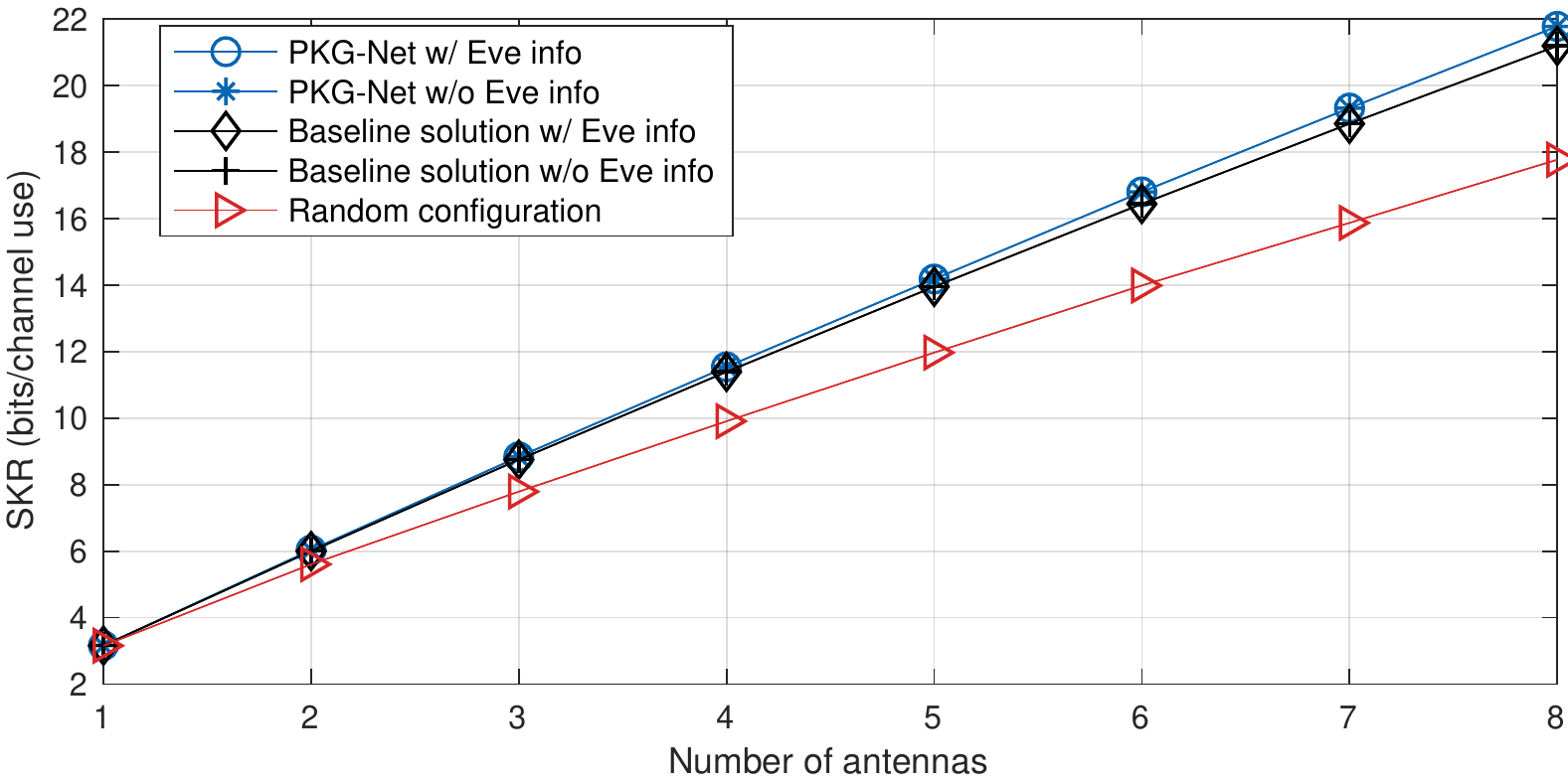}}
\caption{SKR versus $M$ with $P_{\mathrm{max}}=25$ dBm, $\kappa=1$ and $L=36$.}
\label{fig:antenna}
\end{figure}

In Fig. \ref{fig:antenna}, we show the effect of the number of antennas on the SKR. First, it is observed that the SKR  increases linearly with the number of antennas, as more antennas introduce more sub-channels to extract secure keys. Second, we can see that regardless of the number of antennas, the proposed PKG-Net achieves a higher SKR than the two benchmark methods, demonstrating the effectiveness of the proposed DNN-based algorithm. The gap in terms of SKR between the proposed PKG-Net and other benchmark methods increases with the number of antennas, indicating that using more antennas requires a more sophisticated design of BS precoding.  

\begin{figure}[!t]
\centerline{\includegraphics[width=3.4in]{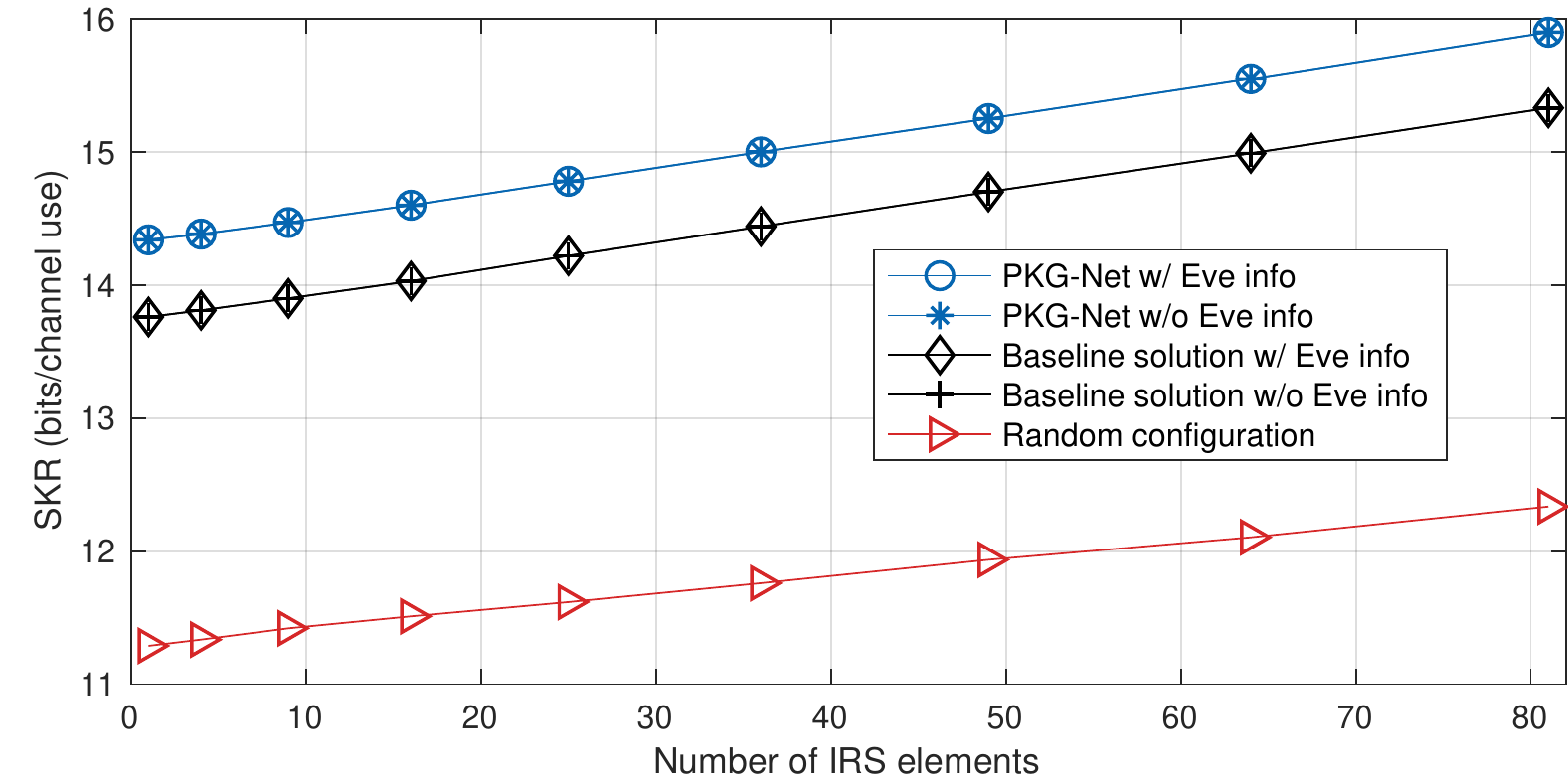}}
\caption{SKR versus $L$ with $P_{\mathrm{max}}=20$ dBm, $\kappa=0$ and $M=6$.}
\label{fig:element}
\end{figure}

Fig. \ref{fig:element} illustrates the effect of the number of IRS elements on the SKR. As can be observed, the SKR monotonically increases with the number of IRS elements, since more IRS elements provide more reflection links that improve the received signals during channel probings. It is noticed that the SKR gain of increasing IRS elements is not as significant as that of increasing BS antennas. This is because in the proposed PKG system, the downlink pilot length is proportional  to the number of antennas. It is expected to achieve a higher SKR when the entire cascaded channel is estimated with a minimum  pilot length of $M(L+1)$, which requires a new design of IRS phase shifting and will be of interest in our future work. 

\begin{figure}[t]
\centerline{\includegraphics[width=3.4in]{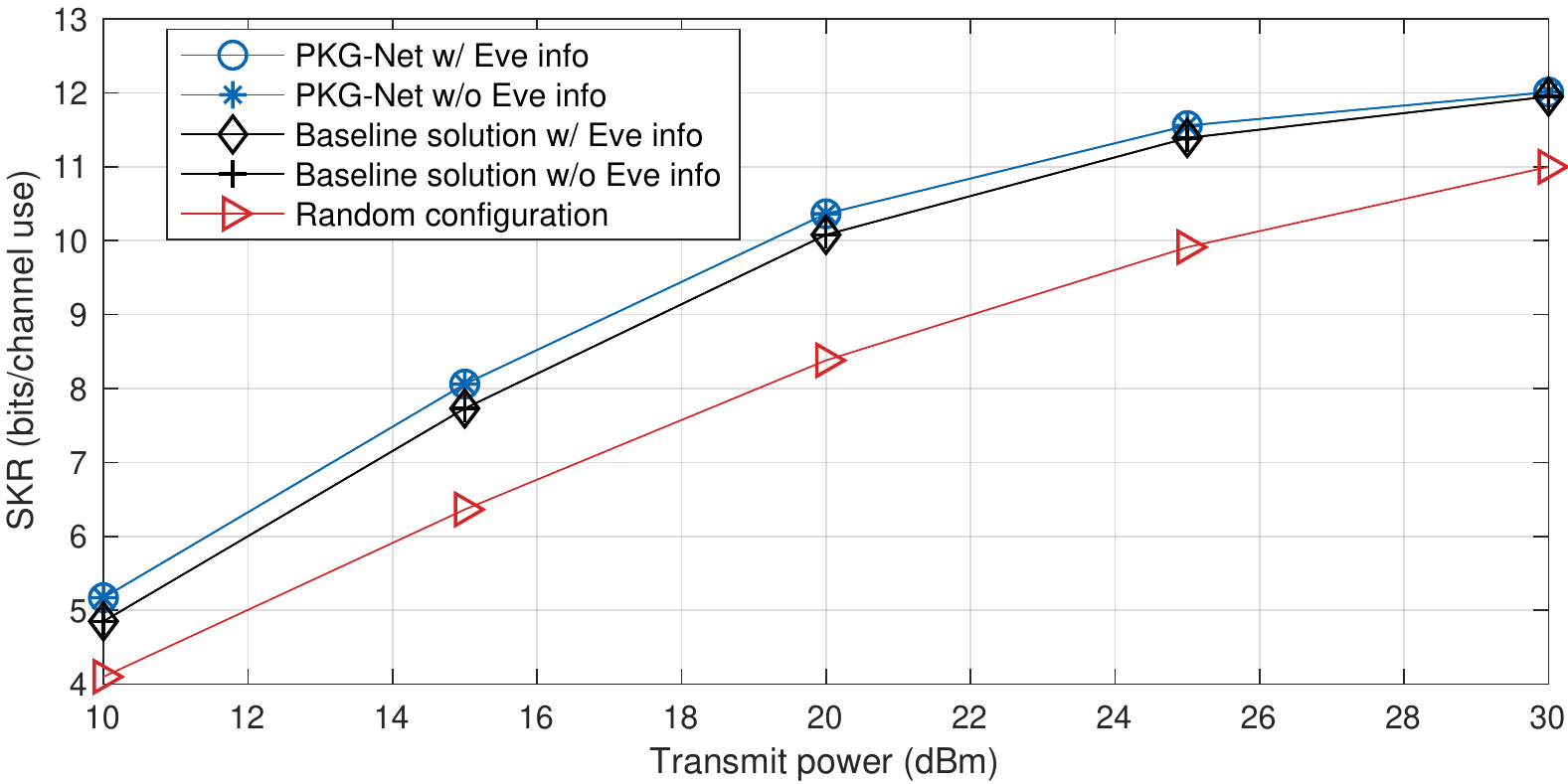}}
\caption{SKR versus $P$ with $\kappa=1$, $M=4$ and $L=36$.}
\label{fig:power}
\end{figure}

In Fig. \ref{fig:power}, the SKR versus the transmit power is displayed. It is shown that the SKR monotonically increases with the transmit power for all the considered optimization methods, which is straightforward due to the fact that a higher transmit power leads to a higher SNR, thereby improving the reciprocity of channel observations at Alice and Bob. 

\begin{figure}[t]
\centerline{\includegraphics[width=3.4in]{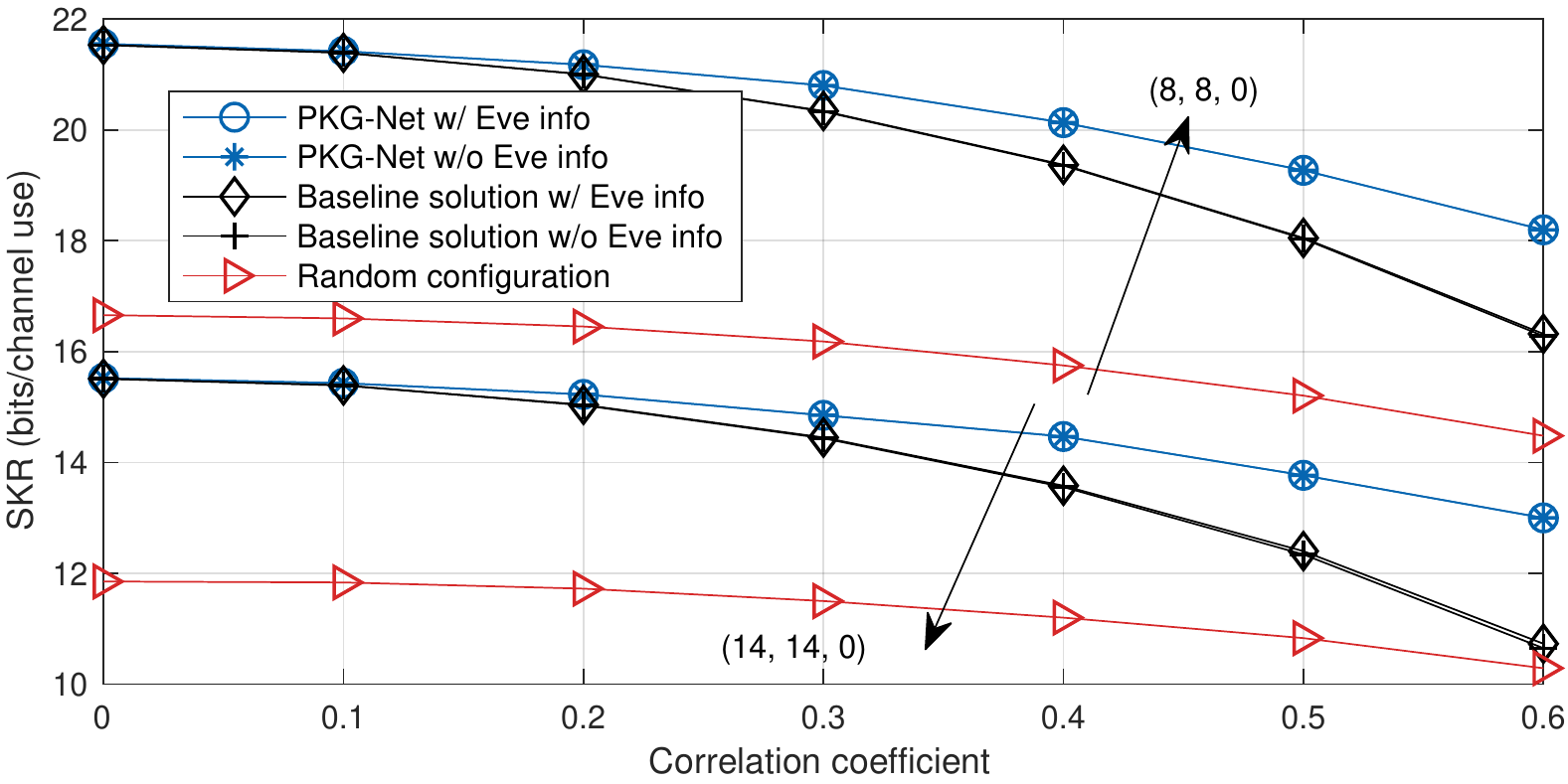}}
\caption{SKR versus $\eta$ with $P_{\mathrm{max}}=20$ dBm, $\kappa=1$, $M=8$ and $L=36$.}
\label{fig:correlation}
\end{figure}

Fig. \ref{fig:correlation} shows the SKR versus the spatial correlation coefficient at the BS for different UE locations. We can see that the SKR monotonically decreases with the correlation coefficient, which illustrates that the existence of spatial correlation degrades the performance of PKG in terms of SKR. In comparison with the benchmark methods, the proposed PKG-Net can effectively design the BS precoding and IRS phase shifting under different spatial correlation conditions. Interestingly, when the correlation coefficient increases, the SKR gain of the proposed PKG-Net becomes more significant compared to the baseline solution. Furthermore, the proposed PKG-Net shows good generalization to UE locations.

\begin{figure}[t]
\centerline{\includegraphics[width=3.4in]{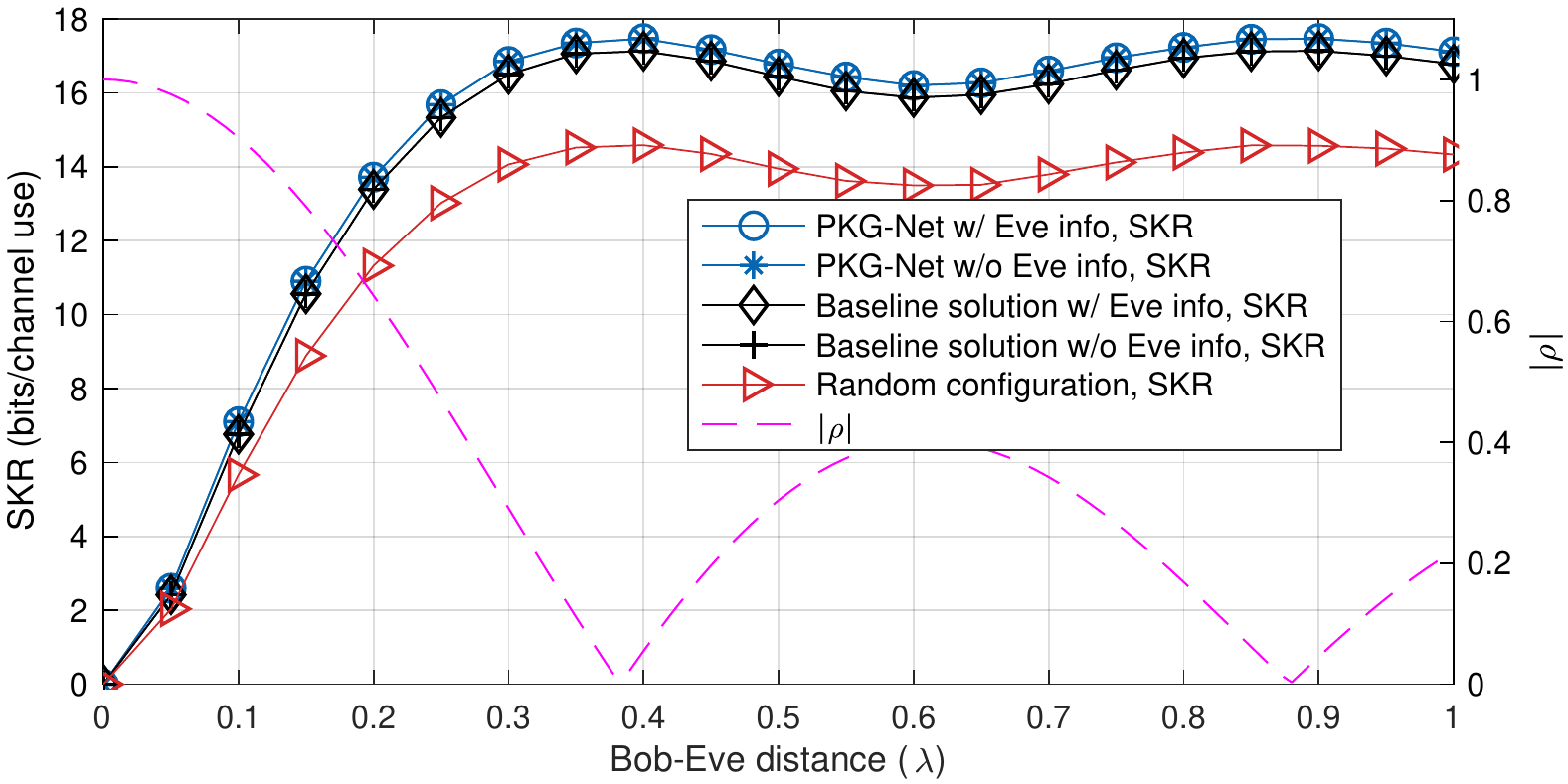}}
\caption{SKR and $|\rho|$ versus the distance between Bob and Eve with $P_{\mathrm{max}}=25$ dBm, $\kappa=1$, $M=6$ and $L=36$.}
\label{fig:distance}
\end{figure}

In Fig. \ref{fig:distance}, we present the effect of the distance between Bob and Eve on the SKR. We can observe that the SKR is inversely proportional to the absolute value of  channel correlation coefficient between Bob and Eve. It is worth noting that a satisfactory SKR can be achieved when the distance between Bob and Eve is more than about 0.3 wavelength. As stated in \cite{Access1}, Bob can easily find Eve if Eve is within half a wavelength of Bob. Hence, the IRS-assisted multi-antenna system can achieve a good PKG performance.

It is interesting to observe that under various system parameters, almost the same SKR can be achieved regardless of whether Eve information is available, indicating that the optimization objectives in case 1 and case 2 have the same solution. Therefore, in practical implementation, we can obtain a satisfactory SKR using only Bob's channel statistics.

{In practice, the conventional optimization algorithms are more suited to run on a classic central processing unit (CPU) architecture, while the matrix multiplication of the DNNs can be efficiently accelerated by a graphics processing unit (GPU). 
As such, it is difficult to make a fair comparison with regard to computational time for different algorithms. However, it is still revealing to show the huge difference in computational time between PKG-Net and the baseline solution using the same hardware configuration \cite{zaher2022learning}. In Table \ref{tab:time}, we present the average computational time of 100 samples for case 1, the UE locations of which are uniformly distributed in the $xy$-plane with $x\in [5, 15]$ and  $y\in [5, 15]$. Other system parameters are set to $P_{\mathrm{max}}=25$ dBm, $\kappa=1$ and $L=36$. Both algorithms run on the same platform, a 4 core Intel Core i7 CPU with 2.2 GHz base frequency and 16GB memory. From Table \ref{tab:time}, it is clear that PKG-Net has a significantly lower computational time than the baseline solution. 
}

\begin{table}[!t]
  \centering
  \caption{Comparison of Running Time for Different Schemes}
  \scalebox{1}{
    \begin{tabular}{|c|C{1.5cm}|C{1.5cm}|C{1.5cm}|}
    \hline
    \textbf{Algorithm} & \textbf{M=2} & \textbf{M=4} & \textbf{M=8} \\
  \hline
    Baseline solution & 74.86 ms & 106 ms & 163 ms\\
  \hline
    PKG-Net & 0.46 ms & 0.91 ms & 2.48 ms \\
    \hline
    \end{tabular}}%
  \label{tab:time}%
\end{table}%

\section{Conclusions}
\label{sec:conslusions}
In this paper, we have studied the PKG in an IRS-assisted multi-antenna system. We have developed a new PKG framework and derived the closed-form expression of SKR considering correlated eavesdropping
channels.  To maximize the SKR, we proposed a novel DNN-based algorithm, PKG-Net, to jointly configure the precoding matrix at the BS  and the phase shift vector at the IRS. The proposed PKG-Net has a good generalization ability to different transmit powers and channel statics. Simulation results show that under various system parameters, the proposed PKG-Net achieves a higher SKR than the benchmark methods. Compared with the baseline solution,  PKG-Net has significantly lower computational complexity and can satisfy real-time processing constraints. Additionally, it is observed that the spatial correlation among BS antennas degrades the PKG performance in terms of SKR, and that the proposed PKG-Net can address the spatial correlation more effectively than other benchmark methods. In our future work, we will extend the proposed PKG system to a multi-user scenario and solve the multi-user precoding problem.

\appendices
\section{Proof of Theorem \ref{theorem1}} \label{AppendixA}
In (\ref{SKR_2}), we further define $R_{\mathrm{sk}}^{1}=I\left(\widetilde{\mathbf{y}}_a;\widetilde{\mathbf{y}}_b^{T}\right)- I\left(\widetilde{\mathbf{y}}_a;\widetilde{\mathbf{y}}_{e}^{T}\right)$ and $R_{\mathrm{sk}}^{2}=I\left(\widetilde{\mathbf{y}}_a;\widetilde{\mathbf{y}}_b^{T}\right)- I\left(\widetilde{\mathbf{y}}_b;\widetilde{\mathbf{y}}_{e}^{T}\right)$. Following~\cite{TWC1}, we have
\begin{align} \label{I_ab3}
I\left(\widetilde{\mathbf{y}}_a;\widetilde{\mathbf{y}}_b^{T}\right) & =H\left(\widetilde{\mathbf{y}}_a\right)+H\left(\widetilde{\mathbf{y}}_b^{T}\right)-H\left(\widetilde{\mathbf{y}}_a,\widetilde{\mathbf{y}}_b^{T}\right) \nonumber \\
& = \log_2 \left(\frac{|\mathbf{R}_{a}||\mathbf{R}_{b}|}{\left|\bm{\mathcal{R}}_{a,b}\right|}\right),
\end{align}
where 
\begin{align} \label{eq:Ra}
\mathbf{R}_{a}&= \mathbb{E}\left\{\widetilde{\mathbf{y}}_{a} \widetilde{\mathbf{y}}_{a}^H\right\}= P_{b}\left(\widetilde{\bm{\theta}}\otimes\mathbf{P}\right)^T\mathbf{R}_{\mathrm{c}}^{\mathrm{U}}\left(\widetilde{\bm{\theta}}\otimes\mathbf{P}\right)^* \!+\! \delta^{2}\mathbf{P}^{T}\mathbf{P}^{*},
\\
\label{eq:Rb}
\mathbf{R}_{b}&=\mathbb{E}\left\{\widetilde{\mathbf{y}}_{b}^{T} \widetilde{\mathbf{y}}_{b}^{*}\right\}=\left(\widetilde{\bm{\theta}}\otimes\mathbf{P}\right)^T\mathbf{R}_{\mathrm{c}}^{\mathrm{U}}\left(\widetilde{\bm{\theta}}\otimes\mathbf{P}\right)^*+\delta^{2}\mathbf{I}_{M},
\end{align}
and
\begin{align} 
\bm{\mathcal{R}}_{a,b}=\begin{bmatrix} \mathbf{R}_{a} &\mathbf{R}_{a,b} \\
    \mathbf{R}_{b,a} &\mathbf{R}_{b}
  \end{bmatrix},
\end{align}
with
\begin{align} 
\mathbf{R}_{\mathrm{c}}^{\mathrm{U}} &= \mathbb{E}\left\{\mathbf{h}_{\mathrm{c}}^{ab}\left(\mathbf{h}_{\mathrm{c}}^{ab}\right)^H\right\} \nonumber \\
&= \begin{bmatrix} \beta_{\mathrm{h}_{ab}}\mathbf{R}_{\mathrm{B}} & \mathbf{0}^{T} \\
    \mathbf{0} & \frac{\beta_{\mathrm{G}}\beta_{\mathrm{f}_{b}}}{\left({1+\kappa}\right)\left({1+\kappa}\right)}\mathbf{R}_{\mathrm{I}}\odot \mathbf{R}_{\mathrm{I}}\otimes \mathbf{R}_{\mathrm{B}}
  \end{bmatrix},           \\
  \label{eq:Rab}
\mathbf{R}_{a, b}&=\mathbf{R}_{b, a}=\mathbb{E}\left\{\widetilde{\mathbf{y}}_{a} \widetilde{\mathbf{y}}_{b}^{*}\right\}= \sqrt{P_{b}}\left(\widetilde{\bm{\theta}}\otimes\mathbf{P}\right)^T\mathbf{R}_{\mathrm{c}}^{\mathrm{U}}\left(\widetilde{\bm{\theta}}\otimes\mathbf{P}\right)^*.
\end{align}
Following the similar steps, we have
\begin{align}  \label{eq:I_ae}
I\left(\widetilde{\mathbf{y}}_a;\widetilde{\mathbf{y}}_e^{T}\right) = \log_2 \left(\frac{|\mathbf{R}_{a}||\mathbf{R}_{e}|}{\left|\bm{\mathcal{R}}_{a,e}\right|}\right),
\end{align}
where
\begin{align} \label{eq:Re}
\mathbf{R}_{e}&=\mathbb{E}\left\{\widetilde{\mathbf{y}}_{e}^{T} \widetilde{\mathbf{y}}_{e}^{*}\right\}=\left(\widetilde{\bm{\theta}}\otimes\mathbf{P}\right)^T\mathbf{R}_{\mathrm{c}}^{\mathrm{E}}\left(\widetilde{\bm{\theta}}\otimes\mathbf{P}\right)^*+\delta^{2}\mathbf{I}_{M},
\end{align}
and 
\begin{align}  \label{eq:Rae}
\bm{\mathcal{R}}_{a,e}=\begin{bmatrix} \mathbf{R}_{a} &\mathbf{R}_{a,e} \\
    \mathbf{R}_{e,a} &\mathbf{R}_{e}
  \end{bmatrix},
\end{align}
with
\begin{align} 
\mathbf{R}_{\mathrm{c}}^{\mathrm{E}} &= \mathbb{E}\left\{\mathbf{h}_{\mathrm{c}}^{ae}\left(\mathbf{h}_{\mathrm{c}}^{ae}\right)^H\right\} \nonumber \\
&= \begin{bmatrix} \beta_{\mathrm{h}_{ae}}\mathbf{R}_{\mathrm{B}} & \mathbf{0}^{T} \\
    \mathbf{0} & \frac{\beta_{\mathrm{G}}\beta_{\mathrm{f}_{e}}}{\left({1+\kappa}\right)\left({1+\kappa}\right)}\mathbf{R}_{\mathrm{I}}\odot \mathbf{R}_{\mathrm{I}}\otimes \mathbf{R}_{\mathrm{B}}
  \end{bmatrix},           \\
\mathbf{R}_{a, e}&=\mathbf{R}_{e, a}=\mathbb{E}\left\{\widetilde{\mathbf{y}}_{a} \widetilde{\mathbf{y}}_{e}^{*}\right\}= \sqrt{P_{b}}\left(\widetilde{\bm{\theta}}\otimes\mathbf{P}\right)^T\mathbf{R}_{\mathrm{c}}^{\mathrm{U,E}}\left(\widetilde{\bm{\theta}}\otimes\mathbf{P}\right)^*,
\end{align}
where 
\begin{align}
&{\mathbf{R}_{\mathrm{c}}^{\mathrm{U,E}}}\!=\!\mathbb{E}\left\{\mathbf{h}_{\mathrm{c}}^{ab} \left(\mathbf{h}_{\mathrm{c}}^{ae}\right)^H\right\}=     \nonumber \\
&\begin{bmatrix} \rho\beta_{\mathrm{h}_{ab}}\beta_{\mathrm{h}_{ae}}\mathbf{R}_{\mathrm{B}} \!&\! \!\mathbf{0}^{T}\! \\
    \mathbf{0} \!&\! \!\frac{\rho\beta_{\mathrm{G}}}{1+\kappa}\sqrt{\frac{\beta_{\mathrm{f}_{b}}\beta_{\mathrm{f}_{e}}}{({1+\kappa})({1+\kappa})}}\mathbf{R}_{\mathrm{I}}\!\odot \!\mathbf{R}_{\mathrm{I}}\!\otimes\! \mathbf{R}_{\mathrm{B}}\!
  \end{bmatrix}\!.\!
\end{align}
Moreover,
\begin{align}   \label{eq:I_be}
I\left(\widetilde{\mathbf{y}}_b;\widetilde{\mathbf{y}}_e^{T}\right) = \log_2 \left(\frac{|\mathbf{R}_{b}||\mathbf{R}_{e}|}{\left|\bm{\mathcal{R}}_{b,e}\right|}\right),
\end{align}
where 
\begin{align} 
\bm{\mathcal{R}}_{b,e}=\begin{bmatrix} \mathbf{R}_{b} &\mathbf{R}_{b,e} \\
    \mathbf{R}_{e,b} &\mathbf{R}_{e}
  \end{bmatrix},
\end{align}
with
\begin{align}  \label{eq:Rbe}
\mathbf{R}_{b, e}\!=\!\mathbf{R}_{e, b}\!=\!\mathbb{E}\left\{\widetilde{\mathbf{y}}_{b} \widetilde{\mathbf{y}}_{e}^{*}\right\}\!=\! \left(\widetilde{\bm{\theta}}\otimes\mathbf{P}\right)^T\mathbf{R}_{\mathrm{c}}^{\mathrm{U,E}}\left(\widetilde{\bm{\theta}}\otimes\mathbf{P}\right)^*.
\end{align}

Then we have 
\begin{align}  \label{eq:Rsk1_2}
&R_{\mathrm{sk}}^{1}=I\left(\widetilde{\mathbf{y}}_a;\widetilde{\mathbf{y}}_b^{T}\right)- I\left(\widetilde{\mathbf{y}}_a;\widetilde{\mathbf{y}}_{e}^{T}\right) = \log_2 \left(\frac{|\mathbf{R}_{b}|\left|\bm{\mathcal{R}}_{a,e}\right|}{|\mathbf{R}_{e}|\left|\bm{\mathcal{R}}_{a,b}\right|}\right), \\
\label{eq:Rsk2_2}
&R_{\mathrm{sk}}^{2}=I\left(\widetilde{\mathbf{y}}_a;\widetilde{\mathbf{y}}_b^{T}\right)- I\left(\widetilde{\mathbf{y}}_b;\widetilde{\mathbf{y}}_{e}^{T}\right) = \log_2 \left(\frac{|\mathbf{R}_{a}|\left|\bm{\mathcal{R}}_{b,e}\right|}{|\mathbf{R}_{e}|\left|\bm{\mathcal{R}}_{a,b}\right|}\right). 
\end{align}

Define that $\mathbf{R}_{\mathrm{Z}}^{\mathrm{U}}=\left(\widetilde{\bm{\theta}}\otimes\mathbf{P}\right)^T\mathbf{R}_{\mathrm{c}}^{\mathrm{U}}\left(\widetilde{\bm{\theta}}\otimes\mathbf{P}\right)^*$, $\mathbf{R}_{\mathrm{Z}}^{\mathrm{E}}=\left(\widetilde{\bm{\theta}}\otimes\mathbf{P}\right)^T\mathbf{R}_{\mathrm{c}}^{\mathrm{E}}\left(\widetilde{\bm{\theta}}\otimes\mathbf{P}\right)^*$ and $\mathbf{R}_{\mathrm{Z}}^{\mathrm{U},\mathrm{E}}=\left(\widetilde{\bm{\theta}}\!\otimes\!\mathbf{P}\right)^T\mathbf{R}_{\mathrm{c}}^{\mathrm{U},\mathrm{E}}\left(\widetilde{\bm{\theta}}\!\otimes\!\mathbf{P}\right)^*$. (\ref{eq:Rsk1}) can be obtained by plugging (\ref{eq:Rb}), (\ref{eq:Rab}), (\ref{eq:Re}) and (\ref{eq:Rae}) into (\ref{eq:Rsk1_2}), and (\ref{eq:Rsk2}) can be obtained by plugging (\ref{eq:Ra}), (\ref{eq:Rab}), (\ref{eq:Re}) and (\ref{eq:Rbe}) into (\ref{eq:Rsk2_2}).
 This completes the proof.

\section{Proof of Lemma \ref{lemma1}} \label{AppendixB}
Plugging (\ref{y_{a}}) and (\ref{y_{b}})
into (\ref{I_ab3}), we have
\begin{align} 
&\mathbf{R}_{a} \nonumber \\
&\!=\! P_{b}\mathbf{P}^{T\!} \mathbb{E}\!\left\{\!\left(\mathbf{h}_{ab}^{\mathrm{NL}} \!+\! \mathbf{G}^{\mathrm{\!NL\!}}\mathbf{\Theta}\mathbf{f}_{b}^{\mathrm{NL}}\right)\left(\left(\mathbf{h}_{ab}^{\mathrm{NL}}\right)^{\!H\!} \!+\! \left(\mathbf{G}^{\mathrm{\!NL\!}}\mathbf{\Theta}\mathbf{f}_{b}^{\mathrm{NL}}\right)^{\!H\!}\right)\!\right\}\!\mathbf{P}^{*\!} \nonumber \\
&~~~~~~~~~~~~~~~~~~~~~~~~~~~~~~~~~~~~~~~~~~~~~~~~~~~~~~~~~+\delta^{2}\mathbf{P}^{T}\mathbf{P}^{*}, \nonumber \\
&\!=\! P_{b}\mathbf{P}^{T} \Big(\mathbb{E}\left\{\left(\mathbf{h}_{ab}^{\mathrm{NL}}\left(\mathbf{h}_{ab}^{\mathrm{NL}}\right)^{\!H\!}\right)\right\}  \nonumber \\
&~~~+
\mathbb{E}\left\{\left(\mathbf{G}^{\mathrm{\!NL\!}}\mathbf{\Theta}\mathbf{f}_{b}^{\mathrm{NL}}\right) \left(\mathbf{G}^{\mathrm{\!NL\!}}\mathbf{\Theta}\mathbf{f}_{b}^{\mathrm{NL}}\right)^{\!H\!}\right\}\Big)\mathbf{P}^{*\!} + \delta^{2}\mathbf{P}^{T}\mathbf{P}^{*}, \nonumber \\
&\!=\! P_{b}\mathbf{P}^{T} \Big(\beta_{\mathrm{h}_{ab}}\mathbf{R}_{\mathrm{B}} \nonumber \\
&~~~+
\mathbb{E}\left\{\left(\mathbf{G}^{\mathrm{\!NL\!}}\mathbf{\Theta}\mathbf{f}_{b}^{\mathrm{NL}}\right) \left(\mathbf{G}^{\mathrm{\!NL\!}}\mathbf{\Theta}\mathbf{f}_{b}^{\mathrm{NL}}\right)^{\!H\!}\right\}\Big)\mathbf{P}^{*\!} + \delta^{2}\mathbf{P}^{T}\mathbf{P}^{*}, 
\end{align}
where 
\begin{align} 
&\mathbb{E}\left\{\left(\mathbf{G}^{\mathrm{\!NL\!}}\mathbf{\Theta}\mathbf{f}_{b}^{\mathrm{NL}}\right) \left(\mathbf{G}^{\mathrm{\!NL\!}}\mathbf{\Theta}\mathbf{f}_{b}^{\mathrm{NL}}\right)^{\!H\!}\right\} \nonumber \\
&= \mathbb{E}\left\{\mathbf{R}_{\mathrm{B}}^{\frac{1}{2}}\widetilde{\mathbf{G}}\mathbf{R}_{\mathrm{I}}^{\frac{1}{2}}\mathbf{\Theta}\mathbf{R}_{\mathrm{I}}^{\frac{1}{2}}\widetilde{\mathbf{f}}_{b}\widetilde{\mathbf{f}}_{b}^{H}\mathbf{R}_{\mathrm{I}}^{\frac{H}{2}}\mathbf{\Theta}^{*}\mathbf{R}_{\mathrm{I}}^{\frac{H}{2}}\widetilde{\mathbf{G}}^{H}\mathbf{R}_{\mathrm{B}}^{\frac{H}{2}}
\right\} \nonumber \\
&\overset{(a)}{=} \mathbf{R}_{\mathrm{B}}\mathbb{E}\left\{\widetilde{\mathbf{G}}\mathbf{R}_{\mathrm{I}}^{\frac{1}{2}}\mathbf{\Theta}\mathbf{R}_{\mathrm{I}}^{\frac{1}{2}}\widetilde{\mathbf{f}}_{b}\widetilde{\mathbf{f}}_{b}^{H}\mathbf{R}_{\mathrm{I}}^{\frac{H}{2}}\mathbf{\Theta}^{*}\mathbf{R}_{\mathrm{I}}^{\frac{H}{2}}\widetilde{\mathbf{G}}^{H}
\right\} \nonumber \\
&=\mathbf{R}_{\mathrm{B}}\mathbb{E}\left\{\text{vec}\left(\widetilde{\mathbf{G}}\mathbf{R}_{\mathrm{I}}^{\frac{1}{2}}\mathbf{\Theta}\mathbf{R}_{\mathrm{I}}^{\frac{1}{2}}\widetilde{\mathbf{f}}_{b}\right)\left(\text{vec}\left(\widetilde{\mathbf{G}}\mathbf{R}_{\mathrm{I}}^{\frac{1}{2}}\mathbf{\Theta}\mathbf{R}_{\mathrm{I}}^{\frac{1}{2}}\widetilde{\mathbf{f}}_{b}\right)\right)^{H}
\right\}  \nonumber \\
&=\mathbf{R}_{\mathrm{B}}\mathbb{E}\bigg\{\left(\left(\mathbf{R}_{\mathrm{I}}^{\frac{1}{2}}\mathbf{\Theta}\mathbf{R}_{\mathrm{I}}^{\frac{1}{2}}\widetilde{\mathbf{f}}_{b}\right)^{\!T\!}\!\otimes\! \mathbf{I}_{M} \right)\text{vec}\left(\widetilde{\mathbf{G}}\right) \left(\text{vec}\left(\widetilde{\mathbf{G}}\right)\right)^{H}      \nonumber \\ 
&~~~~~~~~~~~~~~~~~~~~~~~~~~~~~~~~~~~~~~\times\left(\left(\mathbf{R}_{\mathrm{I}}^{\frac{1}{2}}\mathbf{\Theta}\mathbf{R}_{\mathrm{I}}^{\frac{1}{2}}\widetilde{\mathbf{f}}_{b}\right)^{\!*\!}\!\otimes\! \mathbf{I}_{M} \right)
\bigg\} \nonumber \\
&=\frac{\beta_{\mathrm{G}}}{1+\kappa}\mathbf{R}_{\mathrm{B}}\mathbb{E}\bigg\{\left(\left(\mathbf{R}_{\mathrm{I}}^{\frac{1}{2}}\mathbf{\Theta}\mathbf{R}_{\mathrm{I}}^{\frac{1}{2}}\widetilde{\mathbf{f}}_{b}\right)^{\!T\!}\!\otimes\! \mathbf{I}_{M} \right)
\nonumber \\ 
&~~~~~~~~~~~~~~~~~~~~~~~~~~~~~~~~~~~~~~
\times\left(\left(\mathbf{R}_{\mathrm{I}}^{\frac{1}{2}}\mathbf{\Theta}\mathbf{R}_{\mathrm{I}}^{\frac{1}{2}}\widetilde{\mathbf{f}}_{b}\right)^{\!*\!}\!\otimes\! \mathbf{I}_{M} \right)
\bigg\} \nonumber \\ 
&\overset{(b)}{=} \frac{\beta_{\mathrm{G}}}{1+\kappa}\mathbf{R}_{\mathrm{B}}\mathbb{E}\left\{\left(\mathbf{R}_{\mathrm{I}}^{\frac{1}{2}}\mathbf{\Theta}\mathbf{R}_{\mathrm{I}}^{\frac{1}{2}}\widetilde{\mathbf{f}}_{b}\right)^{\!T\!}\left(\mathbf{R}_{\mathrm{I}}^{\frac{1}{2}}\mathbf{\Theta}\mathbf{R}_{\mathrm{I}}^{\frac{1}{2}}\widetilde{\mathbf{f}}_{b}\right)^{\!*\!}\!\otimes\! \mathbf{I}_{M} 
\right\} \nonumber \\ 
&= \frac{\beta_{\mathrm{G}}}{1+\kappa}\mathbf{R}_{\mathrm{B}}\mathbb{E}\left\{\left(\text{vec}\left(\mathbf{R}_{\mathrm{I}}^{\frac{1}{2}}\mathbf{\Theta}\mathbf{R}_{\mathrm{I}}^{\frac{1}{2}}\widetilde{\mathbf{f}}_{b}\right)\right)^{\!T\!}\!\left(\!\text{vec}\left(\mathbf{R}_{\mathrm{I}}^{\frac{1}{2}}\mathbf{\Theta}\mathbf{R}_{\mathrm{I}}^{\frac{1}{2}}\widetilde{\mathbf{f}}_{b}\right)\right)^{\!*\!} 
\right\} \nonumber \\ 
&= \frac{\beta_{\mathrm{G}}}{1+\kappa}\mathbf{R}_{\mathrm{B}}\mathbb{E}\bigg\{\left(
\text{vec}\left(\mathbf{R}_{\mathrm{I}}^{\frac{1}{2}}\mathbf{\Theta}\mathbf{R}_{\mathrm{I}}^{\frac{1}{2}}\right)\right)^{T}\left(\widetilde{\mathbf{f}}_{b}\otimes \mathbf{I}_{L}\right)      \nonumber \\ 
&~~~~~~~~~~~~~~~~~~~~~~~~~~~~
\times\left(\widetilde{\mathbf{f}}_{b}^{H}\otimes \mathbf{I}_{L}\right)\left(\text{vec}\left(\mathbf{R}_{\mathrm{I}}^{\frac{1}{2}}\mathbf{\Theta}\mathbf{R}_{\mathrm{I}}^{\frac{1}{2}}\right)\right)^{*} 
\bigg\} \nonumber \\ 
&= \frac{\beta_{\mathrm{G}}\beta_{\mathrm{f}_{b}}}{\left({1\!+\!\kappa}\right)\left({1\!+\!\kappa}\right)}\mathbf{R}_{\mathrm{B}}\left(
\text{vec}\!\left(\!\mathbf{R}_{\mathrm{I}}^{\frac{1}{2}}\mathbf{\Theta}\mathbf{R}_{\mathrm{I}}^{\frac{1}{2}}\right)\right)^{\!T\!}    
\!\left(\!\text{vec}\left(\mathbf{R}_{\mathrm{I}}^{\frac{1}{2}}\mathbf{\Theta}\mathbf{R}_{\mathrm{I}}^{\frac{1}{2}}\right)\right)^{\!*\!} \nonumber \\ 
&= \frac{\beta_{\mathrm{G}}\beta_{\mathrm{f}_{b}}}{\left({1\!+\!\kappa}\right)\left({1\!+\!\kappa}\right)}\mathbf{R}_{\mathrm{B}}\bm{\theta}^H(\mathbf{R}_{\mathrm{I}}\odot\mathbf{R}_{\mathrm{I}})\bm{\theta},
\end{align}
where (a) comes from the commutative property of diagonal matrices, and (b) is because $(\mathbf{A}\otimes\mathbf{B})(\mathbf{C}\otimes\mathbf{D})=\mathbf{A}\mathbf{C}\otimes \mathbf{B}\mathbf{D}$. $\mathbf{R}_{b}$ and  $\bm{\mathcal{R}}_{a,b}$ can be obtained following the similar steps, which concludes the proof.

\section{Proof of Corollary \ref{corollary1}} \label{AppendixC}
\begin{align} \label{eq:scalarIab}
\widehat{I}\left(\widetilde{\mathbf{y}}_a;\widetilde{\mathbf{y}}_b^{T}\right)&= \sum_{i=1}^{M}\log_2 \left(\frac{(P_{b}P_{a}\delta_{\mathrm{U}}^{2}p_{i}^{2} + \delta^{2}P_{a})\left(P_{a}\delta_{\mathrm{U}}^{2}p_{i}^{2}+ \delta^{2}\right)}{\delta^{2}P_{b}P_{a}\delta_{\mathrm{U}}^{2}p_{i}^{2}+ \delta^{2}P_{a}^{2}\delta_{\mathrm{U}}^{2}p_{i}^{2} + \delta^{4}P_{a}}\right) \nonumber \\
&= \sum_{i=1}^{M}\log_2 \left(1+ \frac{P_{b}P_{a}\delta_{\mathrm{U}}^{4}p_{i}^{4}}{\delta^{2}P_{b}\delta_{\mathrm{U}}^{2}p_{i}^{2}+ \delta^{2}P_{a}\delta_{\mathrm{U}}^{2}p_{i}^{2} + \delta^{4}}\right).
\end{align}
To find the monotonicity of $\widehat{I}\left(\widetilde{\mathbf{y}}_a;\widetilde{\mathbf{y}}_b^{T}\right)$ with regard to $\delta_{\mathrm{U}}^{2}$, we define $f_{i}(x) = \frac{P_{b}P_{a}x^{2}p_{i}^{4}}{\delta^{2}P_{b}x p_{i}^{2}+ \delta^{2}P_{a}xp_{i}^{2} + \delta^{4}}$. The first derivative of $f_{i}(x)$ is calculated as
\begin{align}
\frac{\mathrm{d}f_{i}(x)}{\mathrm{d}x} = \frac{P_aP_bxp_i^4\left(2\delta^{4} + P_a\delta^{2}xp_i^2 + P_b\delta^{2}xp_i^2\right)}{\left(\delta^{4} + P_a \delta^{2}xp_i^2 + P_b\delta^{2}xp_i^2\right)^2}.
\end{align}
When $x>0$, we have $\frac{\mathrm{d}(f_{i}(x))}{\mathrm{d}x}>0$. Hence, $f_{i}(x)$ monotonically increases with $x$. 
We then rewrite  (\ref{eq:scalarIab}) as 
\begin{align} \label{eq:scalarIab2}
\widehat{I}\left(\widetilde{\mathbf{y}}_a;\widetilde{\mathbf{y}}_b^{T}\right) 
= \sum_{i=1}^{M}\log_2 \left(1+ f_{i}(\delta_{\mathrm{U}}^{2})\right).
\end{align}
From (\ref{eq:scalarIab2}), it is clear that $\widehat{I}\left(\widetilde{\mathbf{y}}_a;\widetilde{\mathbf{y}}_b^{T}\right)$ monotonically increases with  $\delta_{\mathrm{U}}^{2}$. Recall that
\begin{align} 
\delta_{\mathrm{U}}^{2}&=\beta_{\mathrm{h}_{ab}} + \frac{\beta_{\mathrm{G}}\beta_{\mathrm{f}_{b}}}{\left({1+\kappa}\right)\left({1+\kappa}\right)}\bm{\theta}^H(\mathbf{R}_{\mathrm{I}}\odot\mathbf{R}_{\mathrm{I}})\bm{\theta} \nonumber \\
&=\beta_{\mathrm{h}_{ab}} + \frac{\beta_{\mathrm{G}}\beta_{\mathrm{f}_{b}}}{\left({1+\kappa}\right)\left({1+\kappa}\right)}\Bigg(\sum_{l=1}^{L}\left[\mathbf{R}_{\mathrm{I}}\odot\mathbf{R}_{\mathrm{I}}\right]_{l,l}    
+     \nonumber \\
&~~~~~~~~~~~~~~~~\sum_{i=1}^{L}\sum_{j>i}^{L}2\left[\mathbf{R}_{\mathrm{I}}\odot\mathbf{R}_{\mathrm{I}}\right]_{i,j}\text{cos}\big(\theta_{i}-\theta_{j}\big)\Bigg).
\end{align}
As such, $\delta_{\mathrm{U}}^{2}$ is maximized when $\theta_{i}=\theta_{j}, \forall i \ne j$.

\normalem


\begin{thebibliography}{10}
\providecommand{\url}[1]{#1}
\csname url@samestyle\endcsname
\providecommand{\newblock}{\relax}
\providecommand{\bibinfo}[2]{#2}
\providecommand{\BIBentrySTDinterwordspacing}{\spaceskip=0pt\relax}
\providecommand{\BIBentryALTinterwordstretchfactor}{4}
\providecommand{\BIBentryALTinterwordspacing}{\spaceskip=\fontdimen2\font plus
\BIBentryALTinterwordstretchfactor\fontdimen3\font minus
  \fontdimen4\font\relax}
\providecommand{\BIBforeignlanguage}[2]{{%
\expandafter\ifx\csname l@#1\endcsname\relax
\typeout{** WARNING: IEEEtran.bst: No hyphenation pattern has been}%
\typeout{** loaded for the language `#1'. Using the pattern for}%
\typeout{** the default language instead.}%
\else
\language=\csname l@#1\endcsname
\fi
#2}}
\providecommand{\BIBdecl}{\relax}
\BIBdecl

\bibitem{chen2023machine}
C.~Chen, J.~Zhang, T.~Lu, M.~Sandell, and L.~Chen, ``Machine learning-based
  secret key generation for {IRS}-assisted multi-antenna systems,'' \emph{arXiv
  preprint arXiv:2301.08179}, 2023.

\bibitem{Magzine1}
A.~Chorti, A.~N. Barreto, S.~K{\"o}psell, M.~Zoli, M.~Chafii, P.~Sehier,
  G.~Fettweis, and H.~V. Poor, ``Context-aware security for 6g wireless: the
  role of physical layer security,'' \emph{IEEE Commun. Stand. Mag.}, vol.~6,
  no.~1, pp. 102--108, 2022.

\bibitem{nguyen2021security}
V.-L. Nguyen, P.-C. Lin, B.-C. Cheng, R.-H. Hwang, and Y.-D. Lin, ``Security
  and privacy for {6G}: A survey on prospective technologies and challenges,''
  \emph{IEEE Commun. Surv. Tutor.}, vol.~23, no.~4, pp. 2384--2428, 2021.

\bibitem{Access1}
J.~Zhang, G.~Li, A.~Marshall, A.~Hu, and L.~Hanzo, ``A new frontier for iot
  security emerging from three decades of key generation relying on wireless
  channels,'' \emph{IEEE Access}, vol.~8, pp. 138\,406--138\,446, 2020.

\bibitem{waqas2018social}
M.~Waqas, M.~Ahmed, Y.~Li, D.~Jin, and S.~Chen, ``Social-aware secret key
  generation for secure device-to-device communication via trusted and
  non-trusted relays,'' \emph{IEEE Trans. Wirel. Commun.}, vol.~17, no.~6, pp.
  3918--3930, 2018.

\bibitem{bakcsi2019secret}
S.~Bak{\c{s}}i and D.~C. Popescu, ``Secret key generation with precoding and
  role reversal in {MIMO} wireless systems,'' \emph{IEEE Trans. Wirel.
  Commun.}, vol.~18, no.~6, pp. 3104--3112, 2019.

\bibitem{premnath2012secret}
S.~N. Premnath \emph{et~al.}, ``Secret key extraction from wireless signal
  strength in real environments,'' \emph{IEEE Trans. Mobile Comput.}, vol.~12,
  no.~5, pp. 917--930, 2013.

\bibitem{zhang2018channel}
J.~Zhang, A.~Marshall, and L.~Hanzo, ``Channel-envelope differencing eliminates
  secret key correlation: {LoRa}-based key generation in low power wide area
  networks,'' \emph{IEEE Trans. Veh. Technol.}, vol.~67, no.~12, pp.
  12\,462--12\,466, 2018.

\bibitem{chen2022sample}
C.~Chen, ``Sample-grouping-based vector quantization for secret key extraction
  from atmospheric optical wireless channels,'' \emph{IEEE Trans. Wirel.
  Commun.}, vol.~21, no.~11, pp. 8905--8918, 2022.

\bibitem{aldaghri2020physical}
N.~Aldaghri and H.~Mahdavifar, ``Physical layer secret key generation in static
  environments,'' \emph{IEEE Trans. Inf. Forensics Secur.}, vol.~15, pp.
  2692--2705, 2020.

\bibitem{bjornson2022reconfigurable}
E.~Bj{\"o}rnson, H.~Wymeersch, B.~Matthiesen, P.~Popovski, L.~Sanguinetti, and
  E.~de~Carvalho, ``Reconfigurable intelligent surfaces: A signal processing
  perspective with wireless applications,'' \emph{IEEE Signal Process. Mag.},
  vol.~39, no.~2, pp. 135--158, 2022.

\bibitem{wu2021intelligent}
Q.~Wu, S.~Zhang, B.~Zheng, C.~You, and R.~Zhang, ``Intelligent reflecting
  surface-aided wireless communications: A tutorial,'' \emph{IEEE Trans
  Commun.}, vol.~69, no.~5, pp. 3313--3351, 2021.

\bibitem{CL1}
T.~Lu, L.~Chen, J.~Zhang, K.~Cao, and A.~Hu, ``Reconfigurable intelligent
  surface assisted secret key generation in quasi-static environments,''
  \emph{{IEEE} Commun. Lett.}, vol.~26, no.~2, pp. 244--248, Feb. 2022.

\bibitem{staat2021intelligent}
P.~Staat, H.~Elders-Boll, M.~Heinrichs, R.~Kronberger, C.~Zenger, and C.~Paar,
  ``Intelligent reflecting surface-assisted wireless key generation for
  low-entropy environments,'' in \emph{Proc. IEEE 32nd Annual Int. Symp. on
  Personal, Indoor and Mobile Radio Commun. (PIMRC)}, 2021, pp. 745--751.

\bibitem{TVT1}
Z.~Ji, P.~L. Yeoh, D.~Zhang, G.~Chen, Z.~He, H.~Yin, and Y.~li, ``Secret key
  generation for intelligent reflecting surface assisted wireless communication
  networks,'' \emph{{IEEE} Trans. Veh. Technol.}, vol.~70, no.~1, pp.
  1030--1034, Jan. 2021.

\bibitem{TIFS2}
G.~Li, C.~Sun, W.~Xu, M.~D. Renzo, and A.~Hu, ``On maximizing the sum secret
  key rate for reconfigurable intelligent surface-assisted multiuser systems,''
  \emph{{IEEE} Trans. Inf. Forensics Security}, vol.~17, pp. 211--225, Jan.
  2022.

\bibitem{zhang2020prospective}
J.~Zhang, E.~Bj{\"o}rnson, M.~Matthaiou, D.~W.~K. Ng, H.~Yang, and D.~J. Love,
  ``Prospective multiple antenna technologies for beyond {5G},'' \emph{IEEE J.
  Sel. Areas Commun.}, vol.~38, no.~8, pp. 1637--1660, 2020.

\bibitem{chen2022deployment}
C.~Chen, J.~Zhang, X.~Chu, and J.~Zhang, ``On the deployment of small cells in
  {3D} {HetNets} with multi-antenna base stations,'' \emph{IEEE Trans. Wirel.
  Commun.}, vol.~21, no.~11, pp. 9761--9774, 2022.

\bibitem{hu2022reconfigurable}
L.~Hu, G.~Li, X.~Qian, A.~Hu, and D.~W.~K. Ng, ``Reconfigurable intelligent
  surface-assisted secret key generation in spatially correlated channels,''
  \emph{arXiv preprint arXiv:2211.03132}, 2022.

\bibitem{lu2022joint}
T.~Lu, L.~Chen, J.~Zhang, C.~Chen, and A.~Hu, ``Joint precoding and phase shift
  design in reconfigurable intelligent surfaces-assisted secret key
  generation,'' \emph{IEEE Trans. Inf. Forensics Secur.}, early access, Apr.
  20, 2023, doi: 10.1109/TIFS.2023.3268881.

\bibitem{Survey1}
J.~Wang, C.~Jiang, H.~Zhang, Y.~Ren, K.-C. Chen, and L.~Hanzo, ``Thirty years
  of machine learning: The road to pareto-optimal wireless networks,''
  \emph{IEEE Commun. Surv. Tutor.}, vol.~22, no.~3, pp. 1472--1514, 2020.

\bibitem{Magzine2}
N.~Kato, B.~Mao, F.~Tang, Y.~Kawamoto, and J.~Liu, ``Ten challenges in
  advancing machine learning technologies toward {6G},'' \emph{IEEE Wireless
  Commun.}, vol.~27, no.~3, pp. 96--103, 2020.

\bibitem{Access2}
A.~Taha, M.~Alrabeiah, and A.~Alkhateeb, ``Enabling large intelligent surfaces
  with compressive sensing and deep learning,'' \emph{IEEE Access}, vol.~9, pp.
  44\,304--44\,321, 2021.

\bibitem{JSAC1}
C.~Huang, R.~Mo, and C.~Yuen, ``Reconfigurable intelligent surface assisted
  multiuser {MISO} systems exploiting deep reinforcement learning,'' \emph{IEEE
  J. Sel. Areas Commun.}, vol.~38, no.~8, pp. 1839--1850, 2020.

\bibitem{zhang2021deep}
S.~Zhang, S.~Zhang, F.~Gao, J.~Ma, and O.~A. Dobre, ``Deep learning optimized
  sparse antenna activation for reconfigurable intelligent surface assisted
  communication,'' \emph{IEEE Trans. Commun.}, vol.~69, no.~10, pp. 6691--6705,
  2021.

\bibitem{chen2023distributed}
C.~Chen, S.~Xu, J.~Zhang, and J.~Zhang, ``A distributed machine learning-based
  approach for {IRS}-enhanced cell-free {MIMO} networks,'' \emph{arXiv preprint
  arXiv:2301.08077}, 2023.

\bibitem{jiao2021machine}
L.~Jiao, G.~Sun, J.~Le, and K.~Zeng, ``Machine learning-assisted wireless phy
  key generation with reconfigurable intelligent surfaces,'' in \emph{Proc. of
  the 3rd ACM Workshop on Wireless Security and Mach. Learning}, 2021, pp.
  61--66.

\bibitem{WCL1}
E.~Bj{\"o}rnson and L.~Sanguinetti, ``Rayleigh fading modeling and channel
  hardening for reconfigurable intelligent surfaces,'' \emph{{IEEE} Wireless
  Commun. Lett.}, vol.~10, no.~4, pp. 830--834, Apr. 2020.

\bibitem{GLOBECOM1}
Q.~Zhu and Y.~Hua, ``Optimal pilots for maximal capacity of secret key
  generation,'' in \emph{Proc. IEEE GLOBECOM 2019}, Hawaii, USA, Dec. 2019, pp.
  0--5.

\bibitem{TIFS3}
J.~W. Wallace and R.~K. Sharma, ``Automatic secret keys from reciprocal {MIMO}
  wireless channels: Measurement and analysis,'' \emph{{IEEE} Trans. Inf.
  Forensics Security}, vol.~5, no.~3, pp. 381--392, 2010.

\bibitem{TIT1}
U.~Maurer, ``Secret key agreement by public discussion from common
  information,'' \emph{IEEE Trans. Inf. Theory.}, vol.~39, no.~3, pp. 733--742,
  1993.

\bibitem{CL2}
J.~Zhang, B.~He, T.~Q. Duong, and R.~Woods, ``On the key generation from
  correlated wireless channels,'' \emph{IEEE Commun. Lett.}, vol.~21, no.~4,
  pp. 961--964, 2017.

\bibitem{TWC1}
B.~T. Quist and M.~A. Jensen, ``Maximization of the channel-based key
  establishment rate in {MIMO} systems,'' \emph{IEEE Trans. Wirel. Commun.},
  vol.~14, no.~10, pp. 5565--5573, 2015.

\bibitem{TSP1}
C.~Xing, Y.~Jing, S.~Wang, S.~Ma, and H.~V. Poor, ``New viewpoint and
  algorithms for water-filling solutions in wireless communications,''
  \emph{{IEEE} Trans. Signal Process.}, vol.~68, pp. 1618--1634, Feb. 2020.

\bibitem{buzzi2021ris}
S.~Buzzi, C.~D’Andrea, A.~Zappone, M.~Fresia, Y.-P. Zhang, and S.~Feng,
  ``{RIS} configuration, beamformer design, and power control in single-cell
  and multi-cell wireless networks,'' \emph{IEEE Trans. Cogn. Commun. Netw.},
  vol.~7, no.~2, pp. 398--411, 2021.

\bibitem{zaher2022learning}
M.~Zaher, {\"O}.~T. Demir, E.~Bj{\"o}rnson, and M.~Petrova, ``Learning-based
  downlink power allocation in cell-free massive {MIMO} systems,'' \emph{IEEE
  Trans. Wirel. Commun.}, vol.~22, no.~1, pp. 174--188, 2022.

\end{thebibliography}
\end{document}